\documentclass[draft]{agujournal2019}

\usepackage{url} 
\usepackage{lineno}
\usepackage{soul}
\usepackage{microtype}
\usepackage[utf8]{inputenc} 
\usepackage[T1]{fontenc}    
\usepackage{url}            
\usepackage{booktabs}       
\usepackage{multirow}
\usepackage{nicefrac}       
\usepackage{tikz}
\usepackage{svg}

\usepackage{array}
\usepackage{graphicx}
\usepackage{paralist}
\usepackage{float}
\usepackage[stable]{footmisc}
\usepackage{tikz}
\usepackage{enumitem}
\usepackage{amsmath}
\usepackage{amssymb}
\usepackage{mathtools}
\usepackage{amsthm}
\usepackage{algorithm}
\usepackage{algorithmic}
\usepackage{amsfonts}       
\usepackage{relsize} 
\usepackage[most]{tcolorbox}
\usepackage{makecell}

\draftfalse

\journalname{JGR: Machine Learning and Computation}

\theoremstyle{plain}

\theoremstyle{definition}

\theoremstyle{remark}

\newcommand{\EE}{\mathbb{E}}

\newcommand{\diff}{\mathrm{d}}

\newcommand{\xp}{\{x_i\}}
\newcommand{\xpi}{\xp_{i=1}^{N}}

\newcommand{\W}{\mathcal{W}}
\newcommand{\Z}{\mathcal{Z}}
\newcommand{\TT}{\mathcal{T}}
\newcommand{\RR}{\mathcal{R}}
\newcommand{\tst}{{\text{test}}}
\newcommand{\gen}[1]{#1^{\text{gen}}}

\DeclareMathOperator*{\SW}{SW}

\begin{document}

\title{Generative Modeling of Aerosol State Representations}




\authors{
	Ehsan Saleh\affil{1,4},
	Saba Ghaffari\affil{1,4},
	Jeffrey~H. Curtis\affil{3},
	Lekha Patel\affil{5},
	Peter~A. Bosler\affil{5},
	Nicole Riemer\affil{3},
	Matthew West\affil{2}
}

\affiliation{1}{Department of Computer Science, University of Illinois Urbana-Champaign}
\affiliation{2}{Department of Mechanical Science and Engineering, University of Illinois Urbana-Champaign}
\affiliation{3}{Department of Climate, Meteorology and Atmospheric Sciences, University of Illinois Urbana-Champaign}
\affiliation{4}{National Center for Supercomputing Applications, University of Illinois Urbana-Champaign}
\affiliation{5}{Center for Computing Research, Sandia National Laboratories}

\correspondingauthor{Nicole Riemer}{nriemer@illinois.edu}


\begin{keypoints}
\item \textit{Dimensionality reduction for aerosols}: The study uses variational autoencoders to compress detailed aerosol measurements from hundreds of variables down to just a few, while still preserving important climate-related information.
\item \textit{Performance differences across diagnostics}: The model reconstructs cloud droplet-forming properties most accurately, light-scattering properties moderately well, and ice-forming properties with the most difficulty.
\item \textit{New methods for robustness and realism}: The work introduces a noise-resilient preprocessing strategy and a realism metric based on sliced Wasserstein distance to improve the quality and reliability of generated aerosol data.
\end{keypoints}

\begin{abstract}
Aerosol--cloud--radiation interactions remain among the most uncertain components 
of the Earth's climate system, in part due to the high dimensionality of aerosol 
state representations and the difficulty of obtaining complete \textit{in situ} 
measurements. Addressing these challenges requires methods that distill complex 
aerosol properties into compact yet physically meaningful forms. Generative 
autoencoder models provide such a pathway.  We present a framework for learning deep variational autoencoder (VAE) models 
of speciated mass and number concentration distributions, which capture detailed 
aerosol size--composition characteristics. By compressing hundreds of original 
dimensions into ten latent variables, the approach enables efficient storage 
and processing while preserving the fidelity of key diagnostics, including cloud 
condensation nuclei (CCN) spectra, optical scattering and absorption coefficients, 
and ice nucleation properties.  
Results show that CCN spectra are easiest to reconstruct accurately, optical 
properties are moderately difficult, and ice nucleation properties are the most 
challenging. To improve performance, we introduce a preprocessing optimization 
strategy that avoids repeated retraining and yields latent representations resilient 
to high-magnitude Gaussian noise, boosting accuracy for CCN spectra, optical 
coefficients, and frozen fraction spectra.  
Finally, we propose a novel realism metric---based on the sliced Wasserstein 
distance between generated samples and a held-out test set---for optimizing the 
KL divergence weight in VAEs. Together, these contributions enable compact, robust, 
and physically meaningful representations of aerosol states for large-scale climate 
applications.
\end{abstract}

\section*{Plain Language Summary}

Airborne particles, called aerosols, affect how clouds form, how energy moves through the atmosphere, and the Earth’s climate. Scientists often describe these particles in very high detail, using hundreds of numbers to capture their sizes, types, and how they interact with clouds and with light. While this detail is valuable, it is also difficult to store, share, and use in large climate studies.
In this work, we use a type of computer model that can “compress” this complex aerosol information into just a few numbers while still keeping the important scientific details. This makes it easier and faster to run climate simulations. We also test ways to make the model work well even when the original measurements are noisy or incomplete.
We found that the model is especially good at predicting how aerosols form cloud droplets, somewhat less accurate for how they interact with light, and most challenging for how they help ice form in clouds. To improve reliability, we developed a new method to check that the model’s results stay realistic when compared to real-world data.
Our approach can make climate research more efficient and robust, helping scientists better understand how tiny airborne particles shape weather and long-term climate.

\section{Introduction}

Atmospheric aerosols play a critical role in the Earth's climate system, influencing the planet's radiative balance and cloud properties, yet they remain one of the largest sources of uncertainty in climate projections~\cite{IPCC2021}. Their complex nature, characterized by a multitude of evolving physical and chemical properties such as size, shape, and composition, results in a high-dimensional state representation~\cite{Poeschl2005}. This high dimensionality poses significant challenges for their inclusion in large-scale climate models, particularly concerning computational and storage efficiency, hindering accurate climate simulations~\cite{Riemer2019}.

Aerosol models used in climate and air quality studies already rely on reduced representations, most commonly modal or sectional schemes, to make simulations computationally feasible~\cite{Whitby1997,Jacobson2005}. Modal models, for example, describe the aerosol population with a small number of lognormal modes, each representing a group of particles with similar sizes and compositions~\cite{Vignati2004,Binkowski1995,Bauer2008}. While efficient, these hand-crafted representations impose strong assumptions about size distribution shape, internal versus external mixing, and the evolution of chemical composition. Such assumptions can limit accuracy and flexibility, particularly when modeling processes that depend sensitively on detailed particle mixing state~\cite{Chung2005,Mcfiggans2006,Jacobson2001,Zaveri2010,Fierce2016,Ching2017,yao2022quantifying,zheng2021estimating}.

In contrast, data-driven approaches such as generative models can learn low-dimensional representations directly from particle-resolved simulations without prescribing distribution shapes or mixing assumptions. This allows for compact state spaces that retain physical fidelity while avoiding the rigid constraints of traditional reduced representations.

To address this challenge, we propose a generative modeling framework based on variational autoencoders (VAEs)~\cite{kingma2013auto}. VAEs are designed to compress complex, high-dimensional data into compact latent representations while preserving essential information, making them well suited for aerosol applications. We focus on learning compressed representations of detailed aerosol states, specifically speciated mass and number concentration distributions. The VAEs are trained to encode the high-dimensional aerosol data into a low-dimensional latent space, from which the original data can be reconstructed. To evaluate the fidelity of the learned representations, we assess the accuracy of several key climate-relevant aerosol diagnostics derived from the reconstructed data. These diagnostics include cloud condensation nuclei (CCN) spectra, optical scattering and absorption coefficients, and ice nucleation properties.

\begin{figure*}[t]
	\centering
	\includegraphics[width=0.98\linewidth]{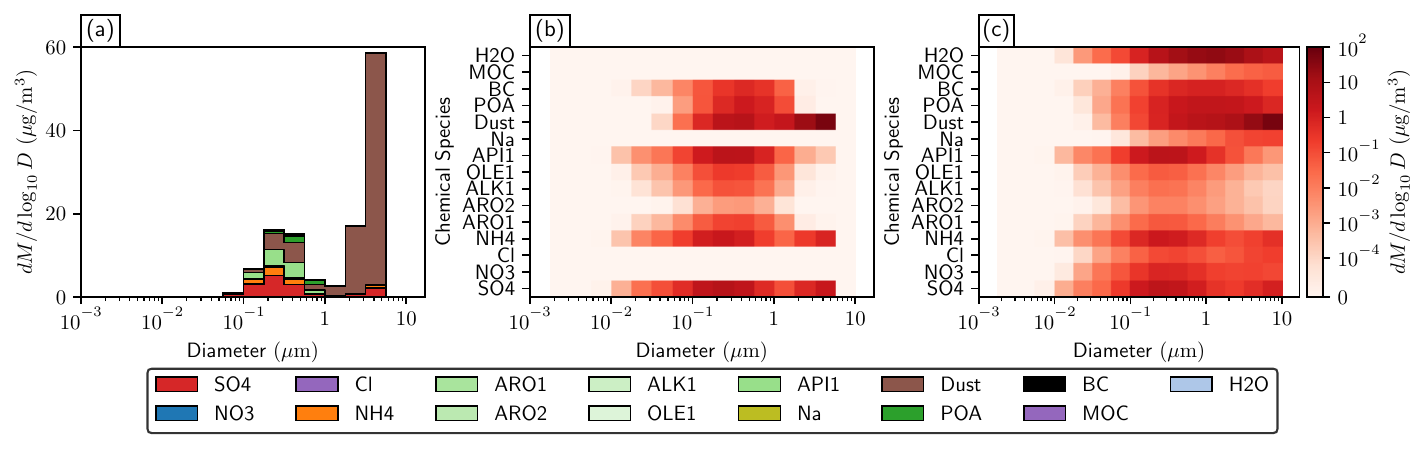}
	\caption{Example of a speciated mass distribution sample. 
	(a) Stacked bar plot of the mass distribution sample, $dM_a/d \log_{10} D$. 
	(b) The same mass distribution but as a heat map. 
	(c) Speciated mass distribution, averaged over all 25000 samples in the dataset.}
	\label{fig:01dataintro}
\end{figure*}

The choice of a generative model was guided by the primary objective of data compression rather than generation. Variational autoencoders (VAEs) are particularly well suited for this task as they excel at encoding high-dimensional data into a compact latent space~\cite{kingma2013auto}. VAEs can be viewed as a nonlinear extension of traditional dimensionality reduction methods like principal component analysis (PCA)~\cite{pearson1901liii} and non-negative matrix factorization (NMF)~\cite{lee1999learning}, offering more flexibility in capturing complex data structures. Their application for data compression and representation learning has been explored in various domains, including atmospheric sciences~\cite{ferracina2025learning}. While other powerful generative models exist, such as generative adversarial networks (GANs)~\cite{goodfellow2014generative}, flow-based models~\cite{rezende2015variational, lipman2022flow}, and diffusion models~\cite{ho2020denoising}, VAEs were selected for their inherent focus on encoding and their training stability.

This study makes several contributions to the representation of aerosol data. First, our results demonstrate that high-dimensional aerosol states, originally comprising hundreds of variables, can be effectively compressed into a latent space of ten dimensions with minimal loss of accuracy in key aerosol diagnostics. This level of compression offers significant memory footprint reduction, which is highly beneficial for large-scale simulation studies. Second, we systematically evaluate the reconstruction performance across different aerosol properties and find that cloud condensation nuclei (CCN) activity is the most accurately reconstructed diagnostic, followed by optical properties, while ice nucleation properties prove to be the most challenging to capture. 
Third, to enhance model robustness, we introduce a computationally efficient pre-processing optimization strategy. This method avoids the need for repeated neural network training by identifying data transformations that are most resilient to noise injection, leading to improved model performance. Consequently, this optimal pre-processing improves the reconstruction of climate-relevant quantities, including CCN spectra, optical scattering and absorption coefficients, and frozen fraction spectra. Finally, we propose a novel realism metric, based on the sliced Wasserstein distance between generated samples and a held-out test set, to provide a principled approach for tuning the Kullback-Leibler (KL) divergence weight in the VAE objective function.

Taken together, these contributions establish a foundation for compact and physically meaningful representations of aerosols. Importantly, they also mark a first step toward learning not just how to compress aerosol states, but how their reduced representations evolve over time—an essential capability for developing efficient surrogate models of aerosol microphysics in future climate studies.

\section{Data}

\begin{figure*}[t]
	\centering
	\includegraphics[width=0.98\linewidth]{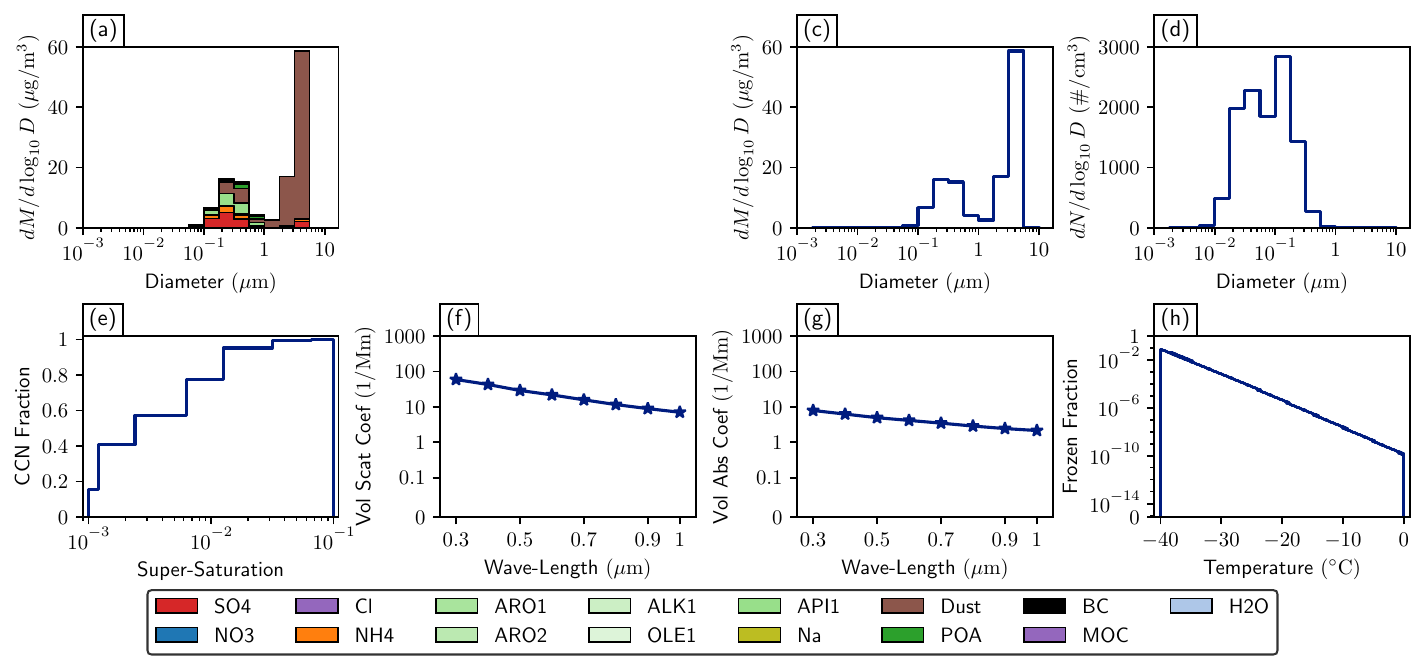}
	\caption{The aerosol diagnostics of the same sample visualized in Figure~\ref{fig:01dataintro}. 
	(a) the speciated mass distribution. 
	(c) the total mass distribution. 
	(d) the number distribution. 
	(e) the CCN spectrum (i.e., the cloud condensation nuclei fraction of the particles at each super-saturation level of critical relative humidity).
	(f) the volume scattering coefficient spectrum. 
	(g) the volume absorption coefficient spectrum.
	(h) the frozen fraction spectrum.}
	\label{fig:02diagintro}
\end{figure*}

This section details the dataset used for training and evaluating our generative models. We begin by describing the source of the data, which is a comprehensive library of aerosol scenarios generated by a particle-resolved model. We then define the specific aerosol state representation used as input to our models, which consists of speciated mass and number distributions. Following this, we outline the calculation of key climate-relevant aerosol diagnostic variables, including CCN spectra, optical properties, and ice nucleation activity. We also describe the methodology for splitting the data into training and testing sets to ensure robust model evaluation.

\subsection{Data Source}

The dataset used in this study is sourced from the scenario library detailed in~\citeA{Gasparik2020}. This library was generated using the particle-resolved aerosol box model PartMC-MOSAIC~\cite{Riemer2009,Zaveri2008}. PartMC explicitly tracks the composition and size of thousands of individual computational particles within an evolving population, resolving mixing state and allowing for a detailed representation of aerosol microphysics. Particle coagulation is simulated using a stochastic Monte Carlo approach, while MOSAIC provides the coupled gas- and aerosol-phase chemistry and thermodynamics. Together, this framework captures emissions, coagulation, dilution with the background, and gas–aerosol partitioning, producing a comprehensive dataset of aerosol populations across diverse atmospheric conditions and emission scenarios.

The library comprises 1000 distinct scenarios, each corresponding to a 24-hour simulation with hourly output (25 time snapshots including the initial state). The aerosol populations within these scenarios are described by 15 chemical species yielding particle-resolved ensembles. Unlike conventional bulk or modal representations, this dataset resolves the full evolution of aerosol mixing state, providing a uniquely stringent test for generative modeling.

\begin{figure*}[t]
	\centering
	\includegraphics[width=0.98\linewidth]{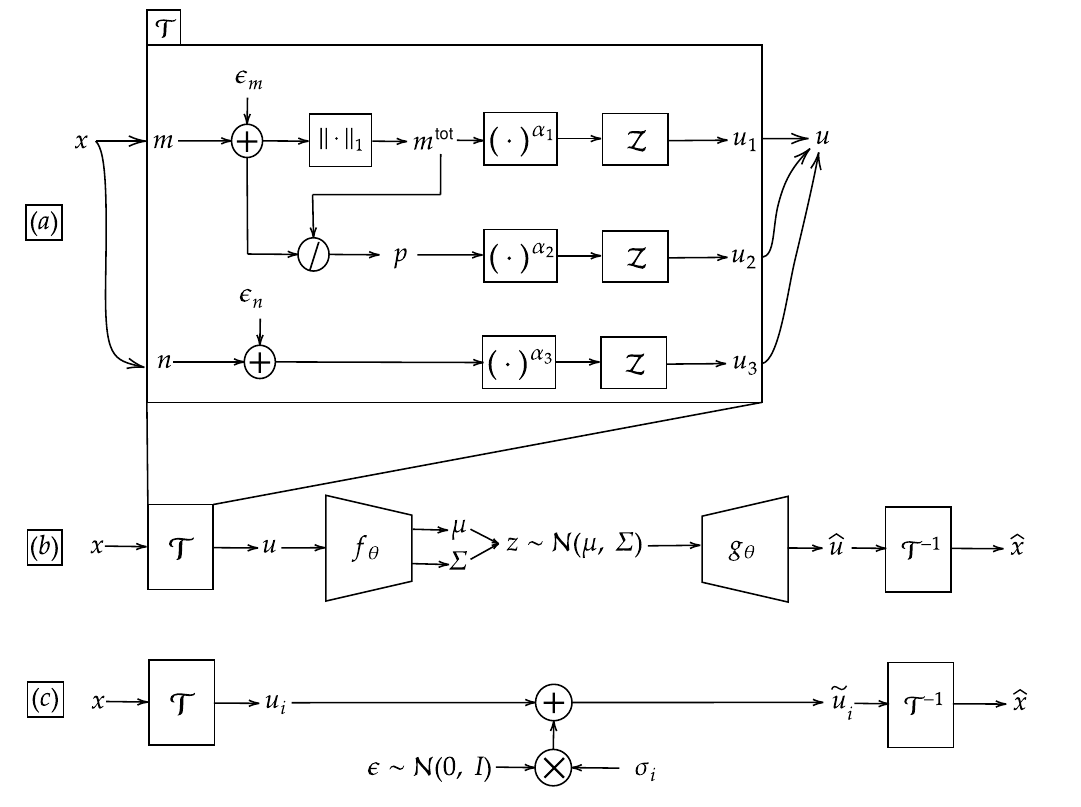}\vspace{-3mm}
	\caption{The modeling pipeline. 
	(a) the preprocessing transformation process. 
	(b) the variational autoencoding pipeline. 
	(c) the preprocessing simulation framework}\label{fig:03schematic}\vspace{-12mm}
\end{figure*}

The 15 chemical species tracked in the model are: Sulfate (SO4), Nitrate (NO3), Chloride (Cl), Ammonium (NH4), Sodium (Na), Dust, Black Carbon (BC), Water (H2O), Primary Organic Aerosol (POA), Marine Organic Compounds (MOC), and five lumped precursors for Secondary Organic Aerosol (SOA): high-yield aromatics (ARO1), low-yield aromatics (ARO2), long-chain alkanes (ALK1), olefins (OLE1), and alpha-pinene (API1). These species encompass primary emissions including dust, POA and BC, and secondary aerosols formed from both inorganic and organic gas-phase precursors.

\subsection{Aerosol State Representation}

Each data sample is represented by the tuple $x=(m, n)$, where $m$ is the
speciated mass distribution and $n$ is the number distribution. The
distributions are discretized across $B=20$ logarithmically spaced diameter bins
ranging from $1\,\mathrm{nm}$ to $10\,\mathrm{\mu m}$. For species $a$ and size
bin $b$, $m_{a,b}$ is the discrete speciated mass size distribution, and $n_b$
is the discrete particle number size distribution. In other words, $m_{a,b}$ and
$n_b$ describe how aerosol mass and number are distributed across particle
sizes. This representation yields a 320-dimensional data vector (20 bins
$\times$ 15 species for mass, plus 20 bins for number).

Figure~\ref{fig:01dataintro} illustrates a sample aerosol population and its corresponding diagnostic variables. Panel (a) displays the speciated mass distribution ($m$), while panel (d) of Figure~\ref{fig:02diagintro} shows the number distribution ($n$). The total mass distribution, derived from summing the speciated mass, is shown in Figure~\ref{fig:02diagintro}(c). The remaining panels of Figure~\ref{fig:02diagintro} present key climate-relevant diagnostics: the cloud condensation nuclei (CCN) fraction spectrum (e), the volume scattering (f) and absorption (g) coefficient spectra, and the frozen fraction spectrum (h).

\begin{figure*}[t]
	\centering
	\includegraphics[width=0.98\linewidth]{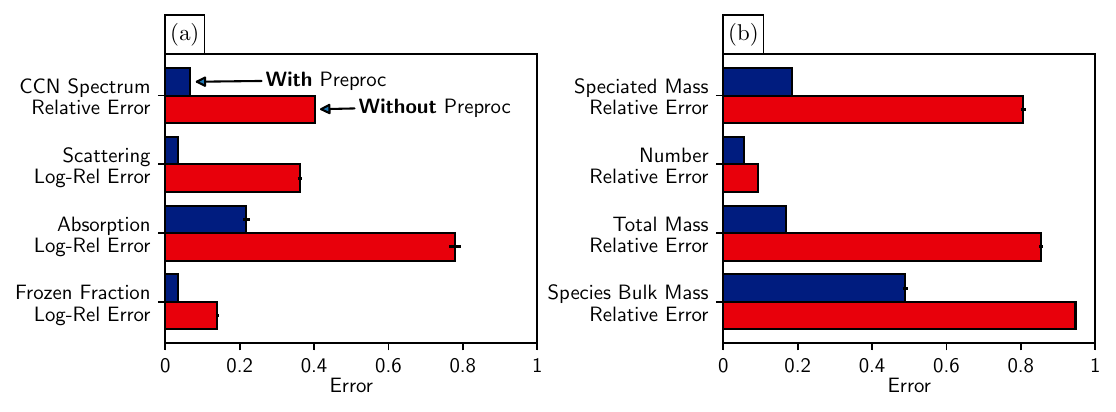}
	\caption{The simulated effect of preprocessing on the aerosol diagnostic metrics. The same process of proportional Gaussian noise injection was applied to both a tuned and a plain preprocessor. (a) the physical aerosol diagnostic metrics. (b) the vector diagnostic metrics.}\label{fig:04simexprmnt}
\end{figure*}

\subsection{Aerosol Diagnostic Variables}

There are four climate relevant diagnostic variables we studied in this paper, CCN spectra, volume absorption and scattering coefficient spectra, and immersion freezing ice nuclei spectra. CCN spectra provide an integrated measure of aerosol size and composition, directly linking particle properties to their ability to form cloud droplets. Because droplet activation is a threshold process that depends on both size and hygroscopicity, CCN spectra serve as a robust benchmark for testing whether compressed representations retain the information most relevant for warm cloud formation. Aerosol scattering and absorption coefficients are central to direct radiative forcing and depend sensitively on mixing state, especially for black carbon and dust. By including optical diagnostics, we directly test whether the latent representations can preserve compositionally dependent absorption and scattering, which are critical for constraining aerosol–radiation interactions. Immersion freezing diagnostics provide a stringent test because ice nucleation is inherently stochastic and often controlled by trace components such as dust and soot. Small reconstruction errors in these components can lead to large differences in frozen fraction spectra. Including this diagnostic therefore probes the limits of compression methods in capturing the rare, nonlinear processes most important for mixed-phase and cirrus cloud formation. The following describes briefly the methods used to calcuate these diagnostics.

\begin{figure*}[t]
	\centering
	\includegraphics[width=0.98\linewidth]{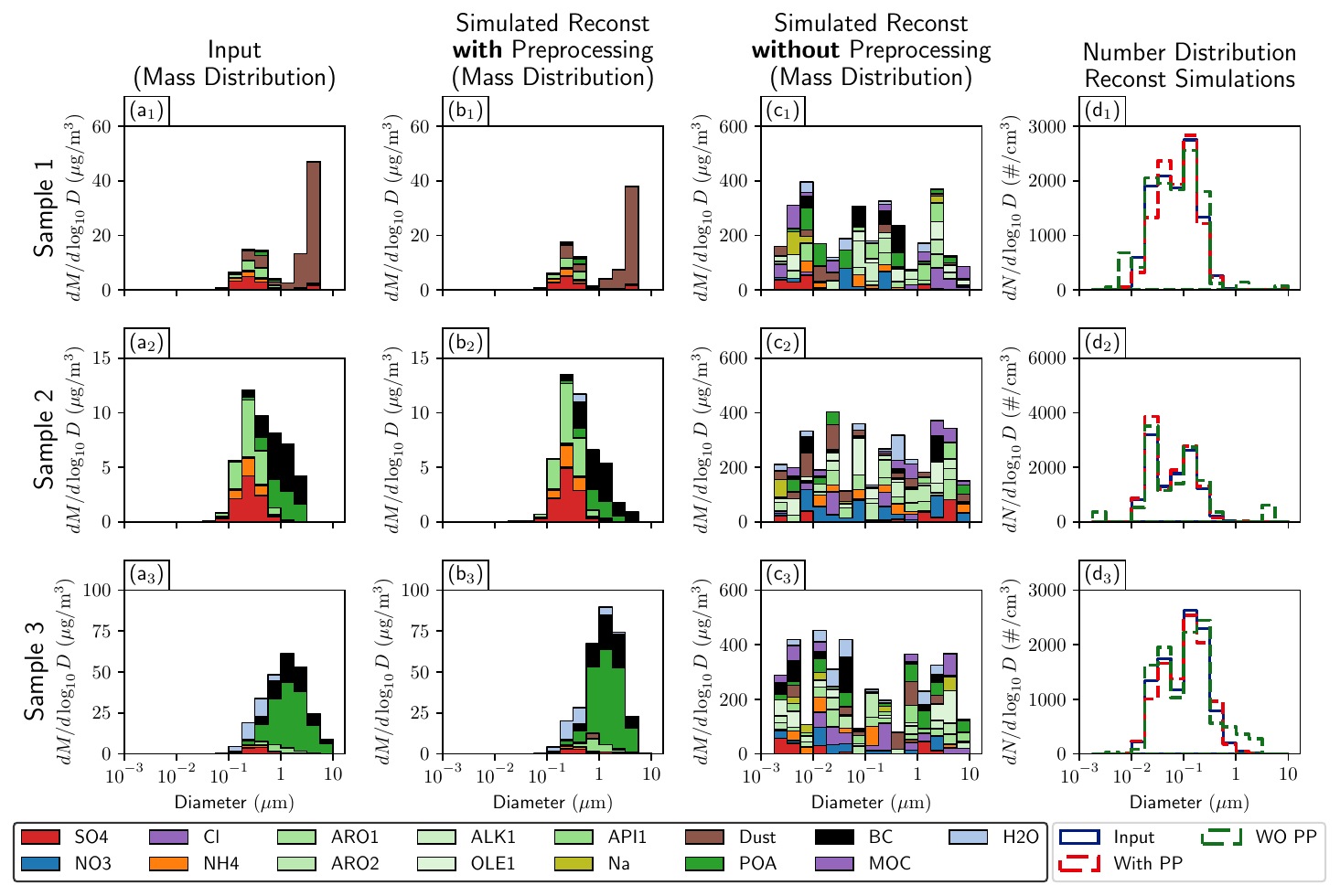}
	\caption{Examples on the effect of optimal vs. plain pre-processing. Each row denotes a single sample. (a$_{1-3}$) the original speciated mass distributions. The (b$_{1-3}$) and (c$_{1-3}$) plots show the noise injected reconstruction $\hat{x}$ for the tuned and plain pre-processors, respectively. (d$_{1-3}$) the number distribution comparison of the original vs. the tuned and plain samples.}\label{fig:05simsample}\vspace{-5mm}
\end{figure*}

\textbf{The CCN Spectrum}: We computed the CCN fraction as in~\citeA{riemer2010}. 
For each bin in the mass and number distribution, the Köhler equation was solved to determine the critical supersaturation based on its size and chemical composition. At a given environmental supersaturation, all bins with critical supersaturations below this threshold were counted as activated, and the resulting activated fraction was used to construct CCN spectra. Figure~\ref{fig:02diagintro}(e) shows the CCN fraction spectrum for the population in Figure~\ref{fig:02diagintro}(a). Because the CCN spectrum is calculated bin by bin, it has a step-like appearance.

\begin{figure*}[!thbp]
	\centering
	\includegraphics[page=1,width=0.98\linewidth]{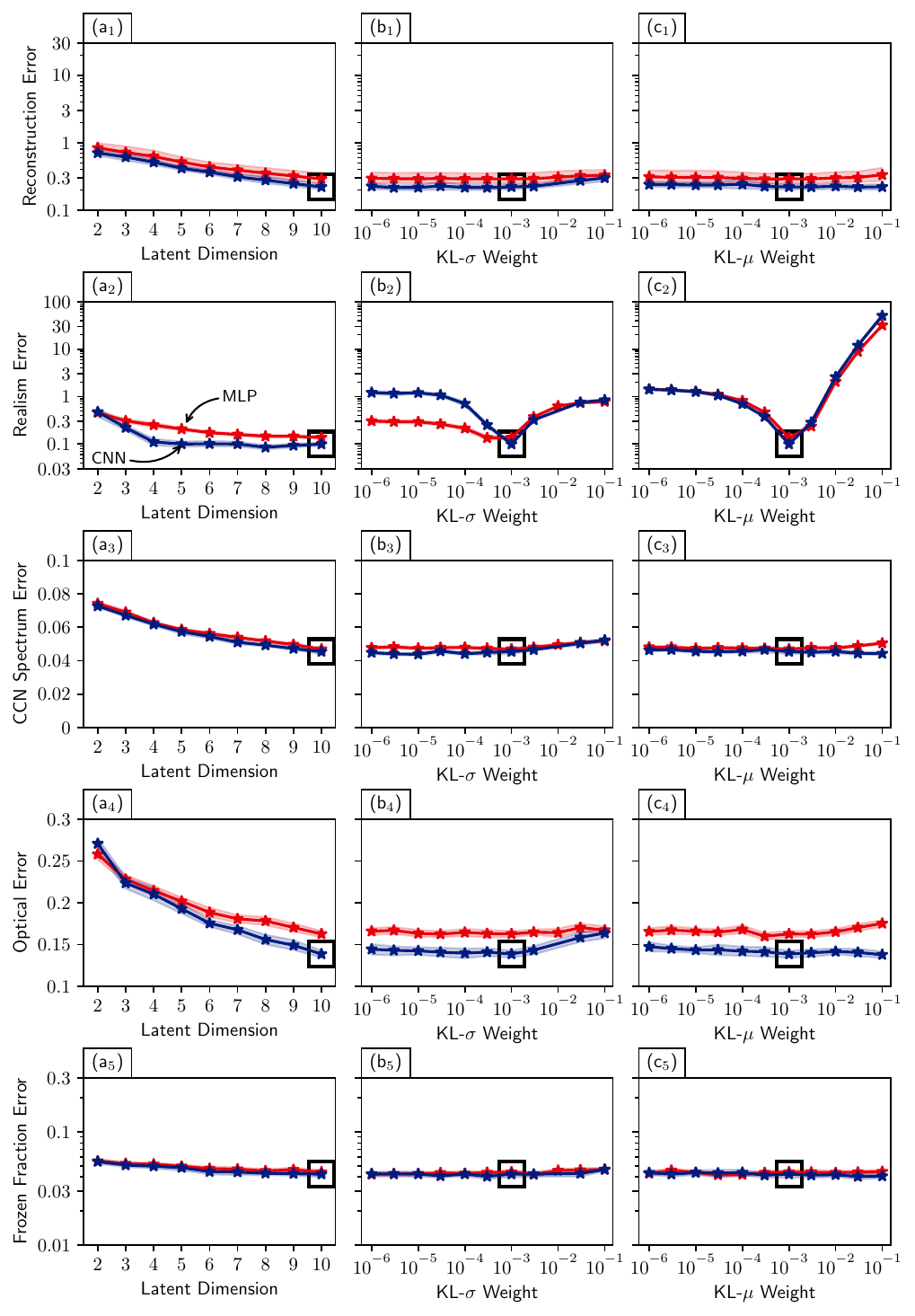}
	\vspace{-3mm}\caption{Ablating the effect of the latent dimensionality and the KL term weights on various performance metrics.}\label{fig:06hpstudy}
\end{figure*}

\textbf{The Volume Absorption and Scattering Coefficient Spectrum}: We computed the volume absorption $\beta_{\rm a}$ and scattering coefficients $\beta_{\rm s}$ following~\citeA{yao2022quantifying} using Mie theory. For bins without dust or black carbon, homogeneous spheres were assumed, with refractive indices determined from composition using volume mixing rules. For dust- and BC-containing bins, a core–shell configuration was assumed, with the absorbing or refractory material treated as the core and the remaining components as the shell. Ensemble optical coefficients were then obtained by summing the bin contributions across the distribution. This treatment captures the influence of both size and composition on aerosol optical behavior. Figure~\ref{fig:02diagintro}(g) shows the $\beta_{\rm a}$ spectrum for the population in Figure~\ref{fig:02diagintro}(a), and Figure~\ref{fig:02diagintro}(f) shows the $\beta_{\rm s}$ spectrum for the population in Figure~\ref{fig:02diagintro}(a).

\textbf{The Frozen Fraction Spectrum}: Ice nucleation properties were evaluated using the ice nucleation active site (INAS) density parameterization~\cite{Hoose2012}. For each bin containing an ice-active component (i.e, dust~\cite{niemand2012particle} or black carbon~\cite{Schill2020}), the number of active sites was determined as a function of particle surface area and temperature. The probability of freezing for each bin was then computed from the product of its surface area and the parameterized active site density. By aggregating over the full distribution, we obtained frozen fraction spectra that represent the immersion freezing behavior of the ensemble. Figure~\ref{fig:02diagintro}(h) shows the frozen fraction spectrum for the population in Figure~\ref{fig:02diagintro}(a).

\subsection{Train and Test Split}

To ensure that our model generalizes to unseen aerosol evolutionary pathways, we partitioned the dataset by randomly splitting entire scenarios into training and testing sets, rather than splitting individual samples. This strategy prevents data leakage from temporally correlated samples within the same scenario. We employed an 80--20 train-test split, assigning 80\% of the scenarios to the training set and 20\% to the test set. To ensure the robustness of our findings, this process was repeated with 10 different randomization seeds. A separate model was trained for each seed, and all statistics reported in this paper were averaged across these 10 randomized runs.

\section{Model}

This section details the generative modeling framework developed to learn compact and robust representations of aerosol states. At the core of our approach is a variational autoencoder (VAE), which we describe first, outlining the architecture of the encoder and decoder networks and the procedures for data reconstruction and generation. A critical component of our framework is a multi-stage preprocessing transformation designed to handle the highly non-Gaussian nature of the input aerosol data. We then introduce a computationally efficient, simulation-based strategy for optimizing the hyperparameters of this transformation. Finally, we describe the iterative procedure used to tune the main VAE hyperparameters, including the latent space dimensionality and the weights of the Kullback–Leibler divergence term in the objective function. Throughout the subsequent sections, we define the relative error between vectors $a$ and $b$ as $\|a-b\|_2/(\|a\|_2 + \|b\|_2)$.

\subsection{Generative Modeling}

We employ a variational autoencoder (VAE) for generative modeling. The process begins with an encoder network, $f_\theta$, which maps a preprocessed input sample, $u$, to a low-dimensional latent representation. Specifically, the encoder outputs the mean ($\mu$) and covariance ($\Sigma$) parameters that define a variational distribution for the sample in the latent space:
\begin{equation}
    \mu, \Sigma = f_\theta(u).
    \label{eq:01enc}
\end{equation}
We utilize a diagonal parameterization for the covariance matrix $\Sigma=\text{diag}(\sigma_1^2, \ldots, \sigma_d^2)$ to simplify the model and reduce computational cost. A latent variable, $z$, is then sampled from this parameterized Normal distribution:
\begin{equation}
    z \sim \mathcal{N}(\mu, \Sigma).
    \label{eq:02latentsample}
\end{equation}

The latent variable $z$ is then passed through the decoder network, $g_{\theta}$, which attempts to reconstruct the original sample. The decoder generates an approximation, $\hat{u}$, of the preprocessed variable $u$:
\begin{equation}
    \hat{u} = g_{\theta}(z).
    \label{eq:03dec}
\end{equation}

To obtain the reconstructed sample, $\hat{x}$, in its original nonprocessed form, we apply the inverse preprocessing transformation, $\TT^{-1}$, to the decoder output, $\hat{u}$:
\begin{equation}
	\hat{x} = \TT^{-1}(\hat{u}).
	\label{eq:04recon}
\end{equation}
Any reconstructed diagnostic variables are then computed from $\hat{x}=(\hat{m}, \hat{n})$.

The model can be trained by minimizing the reconstruction and variational losses:
\begin{equation}\label{eq:04trnloss}
	\mathcal{L} = \EE\Big[\|x - \hat{x}\|_2^2\Big] - \frac{w_{\mu}}{2} \EE\Big[\|\mu\|_2^2\Big] - \frac{w_{\sigma}}{2} \sum_{i=1}^{d} \EE\Big[\sigma_i^2 - 1 + \log(\sigma_i^2)\Big].
\end{equation}
Here, $w_{\mu}$ and $w_{\sigma}$ are weighting terms, and the expectations are replaced with an empirical average over a mini-batch of training samples for stochastic gradient descent.

To generate new samples, the encoder network is not used. Instead, a new latent variable $\gen{z}$ is sampled from a standard normal distribution $\mathcal{N}(0, I)$, and the decoder network is applied to this latent variable to generate a new sample $\gen{x}$:
\begin{equation}
	\gen{z} \sim \mathcal{N}(0, I), \quad \gen{x} = \TT^{-1}(g_{\theta}(\gen{z})).
	\label{eq:05gensample}
\end{equation}

\begin{figure*}[t]
	\centering
	\includegraphics[width=0.98\linewidth]{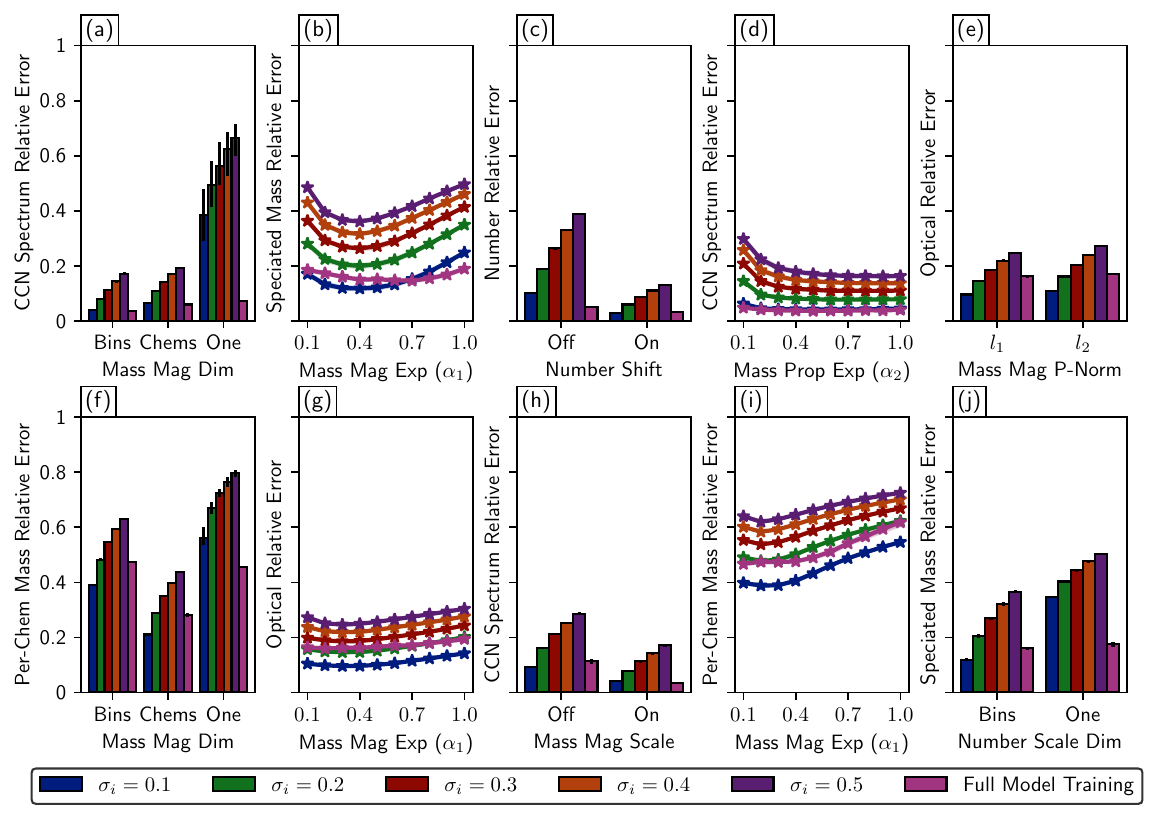}
	\caption{Evaluating the simulation behavior compared to full model training. Five different simulation noise levels corresponding to $\sigma_i\in [0.1, 0.5]$ in Equation~\eqref{eq:14simnoise} were compared against full model trainings. The overall simulation and full training trends appear to match closely.}\label{fig:07simval}
\end{figure*}

\begin{figure*}[!thbp]
	\centering
	\includegraphics[page=1,width=0.98\linewidth]{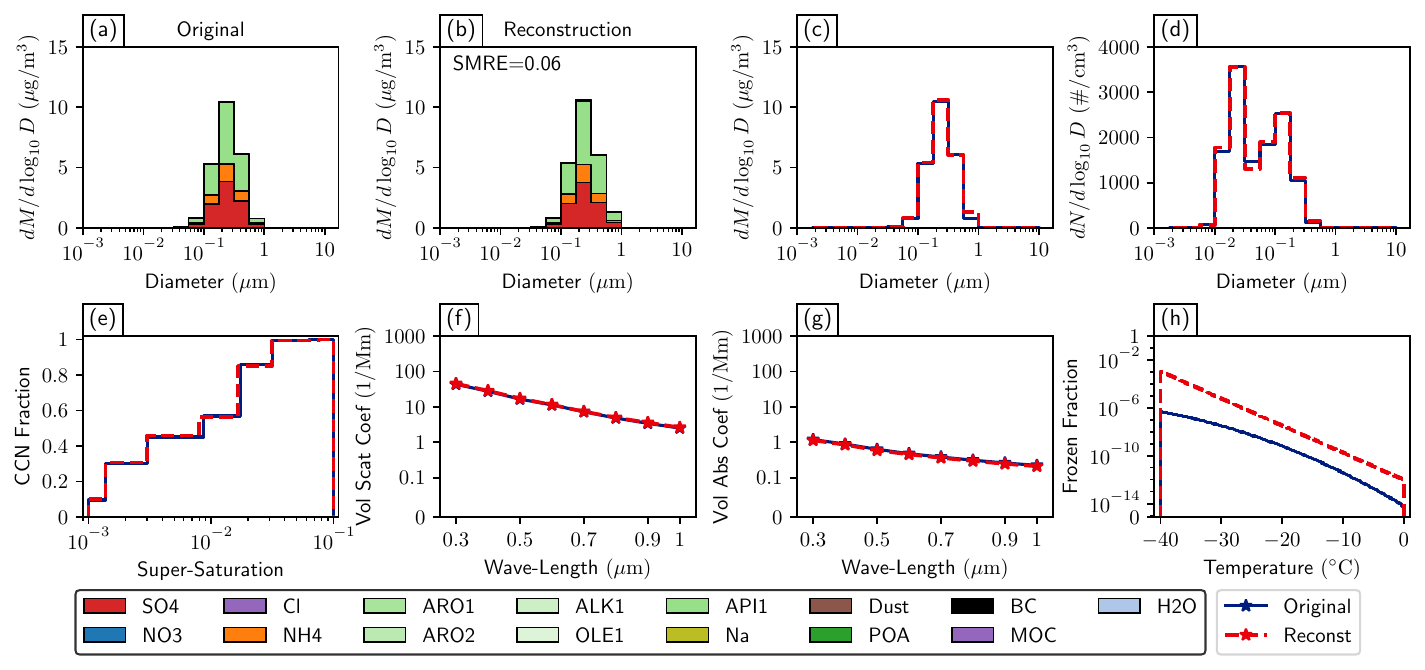}
	\includegraphics[page=2,width=0.98\linewidth]{figures/41_cnn_test_anec.pdf}
	\includegraphics[page=3,width=0.98\linewidth]{figures/41_cnn_test_anec.pdf}
	\vspace{-3mm}\caption{The aerosol diagnostics for the original and reconstructed variants of a particular test sample. The (a)--(h) plots show the speciated mass and number distributions, the CCN spectrum, the volume scattering and absorption coefficient curves, and the frozen fraction spectrum. The (i$_1$)--(i$_{15}$) plots show the mass distributions conditioned for each chemical. The (j$_1$)--(j$_{15}$) plots show the same mass distributions, with the vertical axis being independently scaled. The particular sample in the figure has a speciated mass relative error of 0.06.}\label{fig:08anecdiag}
\end{figure*}

To measure the realism of a population of generated samples, we define a realism metric, $\RR$, as the sliced Wasserstein distance~\cite{kolouri2019gsw} between the distribution of generated samples and the distribution of real samples from a held-out test set:
\begin{equation}
	\RR = \SW(\{x_i^\tst\}_{i=1}^{N}, \{\gen{x}_i\}_{i=1}^{N}),
	\label{eq:06realism}
\end{equation}
where $\{x_i^\tst\}_{i=1}^{N}$ is a set of real samples and $\{\gen{x}_i\}_{i=1}^{N}$ is a set of generated samples. The $\SW$ operator projects the high-dimensional data onto a random 1D direction, namely the $v$ slice, and computes the Wasserstein distance between the projected populations. The average across many slices is called the sliced Wasserstein distance:
\begin{equation}
	\SW(\{x_i^\tst\}_{i=1}^{N}, \{\gen{x}_i\}_{i=1}^{N}) = \EE_{v}[\W(\{v^\top x_i^\tst\}_{i=1}^{N},\, \{v^\top \gen{x}_i\}_{i=1}^{N})].
	\label{eq:07sw}
\end{equation}
The slicing directions $v$ can be sampled uniformly from the unit sphere:
\begin{equation}
	v = \frac{\tilde{v}}{\|\tilde{v}\|}, \quad \text{where} \quad \tilde{v} \sim \mathcal{N}(0, I).
	\label{eq:08slicing}
\end{equation}
However, uniform sampling may cause the metric to be dominated by the principal components of the data, as these directions exhibit the largest variation and contribution to the realism metric. To provide more control, we can sample the slicing directions from a distribution that weights the principal components according to their singular values. This is achieved by sampling from a multivariate normal distribution whose covariance is a function of the singular value decomposition of the data matrix, where an exponent $\alpha$ controls the weighting:
\begin{equation}
	v = \frac{VS^{\alpha}\tilde{v}}{\|VS^{\alpha}\tilde{v}\|}, \quad \text{where} \qquad \tilde{v}\sim \mathcal{N}(0, I).
\end{equation}
Here, $X^\tst_{N\times d} = U_{N\times d} S_{d \times d} V^T_{d\times d}$ is the singular value decomposition of the test data matrix $X^\tst$, and $S^{\alpha}$ is the diagonal matrix of singular values raised to the power of $\alpha$.

A value of $\alpha=0$ corresponds to uniform sampling, $\alpha>0$ emphasizes principal components, and $\alpha<0$ emphasizes non-principal components. Figures~\ref{fig:a01realism} and~\ref{fig:a02realism} in the appendix show the effect of $\alpha$ on the realism metric. For $\alpha \ge 0$, the realism metric is insensitive to latent dimensionality, consistent with a PCA-like behavior where the model prioritizes capturing high-variation components. In contrast, for $\alpha < 0$, higher latent dimensions lead to improved realism, as the model has a greater capacity to encode less variable but still important features of the data distribution.

\subsection{Preprocessing Transformation}

The preprocessing transformation focuses on the speciated mass ($m$) and number ($n$) distributions. These variables are highly non-Gaussian, long tailed, and have a large dynamic range. Figure~\ref{fig:a03qqplot} shows a quantile–quantile plot of the input data with and without preprocessing against the normal distribution, illustrating how the transformation helps to normalize the data distribution and how the unprocessed data is abnormal in distribution.

\begin{figure*}[!thbp]
	\centering
	\includegraphics[width=0.98\linewidth]{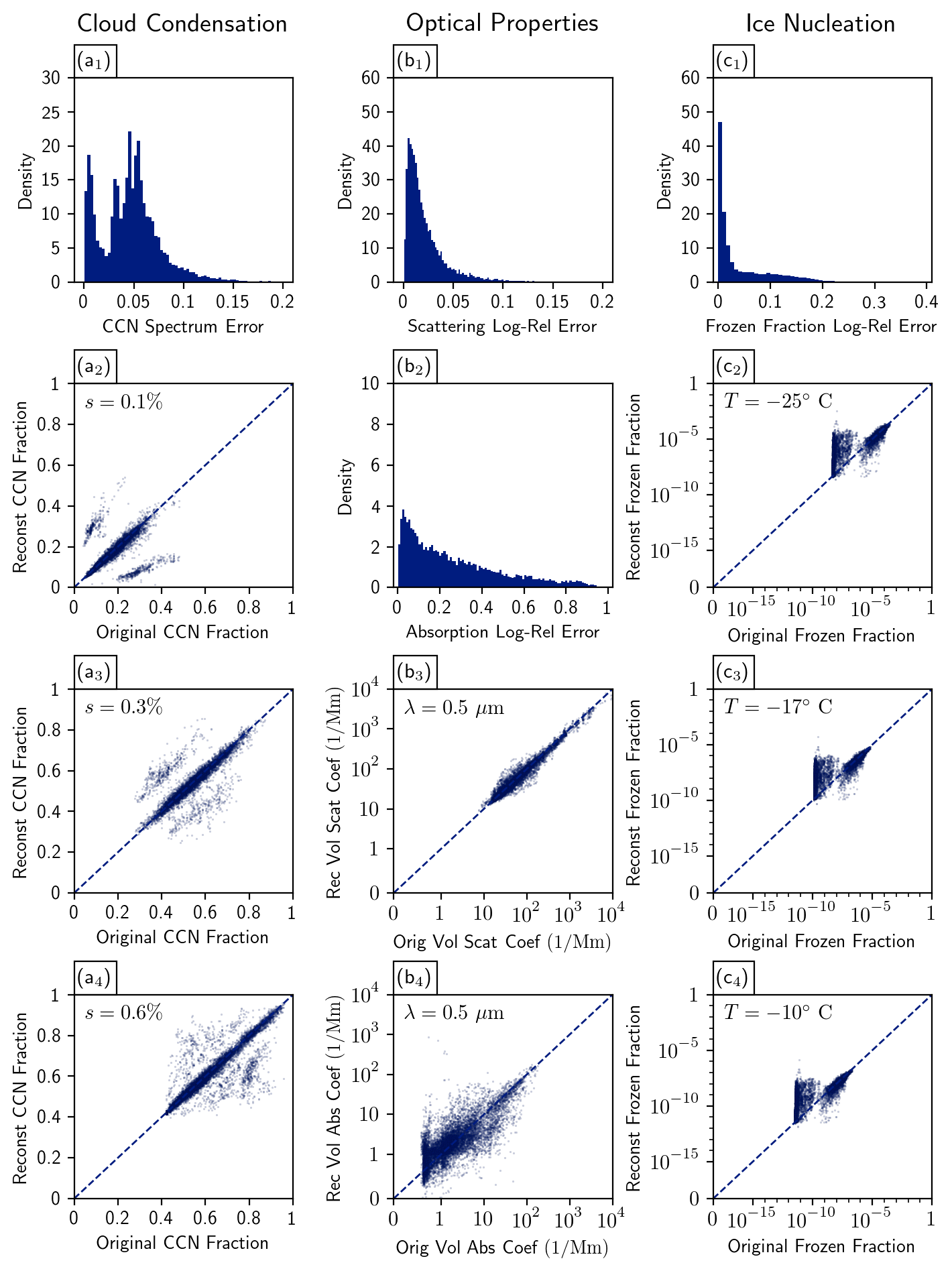}
	\caption{The collective aerosol diagnostic summary plots on the testing portion of the data. 
	The left column shows (1) the histogram of the CCN spectrum error, and (2) the reconstructed vs. original CCN fractions at three different super-saturation levels ($s=0.1\%$, $0.3\%$, and $0.6\%$).
	The middle column shows the scattering and absorption error histograms, and 
	the corresponding reconstructed vs. original scatter plots at a wavelength of $\lambda=0.5 \, {\rm \mu m}$. The right column shows the frozen fraction error histogram, and the reconstructed vs. original scatter plots at three different temperatures of $T=-25 \, {\rm ^{\circ} C}$, $-17 \, {\rm ^{\circ} C}$, and $-10 \, {\rm ^{\circ} C}$.}\label{fig:09smrytest}
\end{figure*}

The preprocessing transformation, $\TT$, maps the input data $x=(m, n)$ to a new representation $u=(u_1, u_2, u_3)$. The components of $u$ are derived from the total mass, mass fractions, and number distribution. First, the total mass in each bin, $m^{\text{tot}}_{b}$, is calculated by summing the mass of each species $a$ in that bin, with a small offset $\epsilon_m$ for numerical stability and taking the absolute value to ensure positivity even if reconstruction generates negative masses:
\begin{equation}
	m^{\text{tot}}_{b} = \sum_{a=1}^{A} |m_{a, b} + \epsilon_m|.
	\label{eq:09mtot}
\end{equation}
This total mass is then transformed using a Box-Cox-like transformation with exponent $\alpha_1$ and standardized to have zero mean and unit variance, yielding $u_1$:
\begin{equation}
	u_1 = \Z\big[\big(m^{\text{tot}}\big)^{\alpha_1}\big].
	\label{eq:10u1}
\end{equation}
Next, the species mass fractions, $p_{a,b}$, are computed by normalizing the mass of each species in a bin by the total mass in that bin:
\begin{equation}
	p_{a,b} = \frac{m_{a,b}}{m^{\text{tot}}_{b}}.
	\label{eq:11p}
\end{equation}
These fractions are also transformed with an exponent $\alpha_2$ and standardized to create $u_2$:
\begin{equation}
	u_2 = \Z[p^{\alpha_2}].
	\label{eq:12u2}
\end{equation}
Finally, the number distribution $n$ is transformed similarly with an exponent $\alpha_3$ and an offset $\epsilon_n$, followed by standardization, to produce $u_3$:
\begin{equation}
	u_3 = \Z\big[\big(n+\epsilon_n\big)^{\alpha_3}\big].
	\label{eq:13u3}
\end{equation}
Figure~\ref{fig:03schematic}(a) summarizes the $\TT$ preprocessing transformation pipeline.

\subsection{Preprocessing Optimization via Simulated Training}

To select the optimal preprocessing transformation, we evaluate the resilience of a given preprocessor $\TT$ to reconstruction errors. We developed a computationally inexpensive performance assay that approximates the full neural model training pipeline (Figure~\ref{fig:03schematic}) with a direct Gaussian noise injection process in the preprocessed space. For a given input $x$, the simulated reconstruction $\tilde{x}$ is obtained by transforming the data, adding noise, and then applying the inverse transformation:
\begin{equation}
	u = \TT(x) = (u_1, u_2, u_3), \qquad \tilde{u}_i = u_i + \sigma_i \cdot \mathcal{N}(0, I), \qquad \tilde{x} = \TT^{-1}(\tilde{u}).
	\label{eq:14simnoise}
\end{equation}
The noise amplitude $\sigma$ is a scalar proportional to the standard deviation of the preprocessed variable $u$ over the training samples. Specifically, we used $\sigma_i = 0.3 \, \sigma_{u_i}$ for each $u_i$ component, where $\sigma_{u_i}$ is the scalar standard deviation of the preprocessed $u_i$ values over the training samples. Figure~\ref{fig:03schematic}(c) summarizes this process.

As shown in Figure~\ref{fig:04simexprmnt}, the simulated reconstruction \textit{with} preprocessing is more accurate than \textit{without} it. Furthermore, Figure~\ref{fig:05simsample} illustrates that optimally preprocessed samples are more resilient to noise injection than unprocessed data, a finding substantiated by the highly non-Gaussian nature of the original data (Figure~\ref{fig:a03qqplot}). While Figure~\ref{fig:a04hpstudy} demonstrates the sensitivity of model performance to key hyperparameters like the Box–Cox exponents, we relied on this simulation framework for tuning, as it efficiently identifies transformations that are robust to error. For instance, unit exponents, which approximate an identity transformation, result in abnormal data distributions and degrade the reconstruction of optical and ice nucleation properties. Figure~\ref{fig:07simval} shows that the overall trends of the simulation framework closely follow the full model training trends.

\subsection{Hyper-Parameter Optimization}

The hyper-parameter optimization process began with tuning and fixing the pre-processing parameters. This step is crucial because the choice of pre-processing can unfairly manipulate reconstruction error metrics; for instance, scaling the data by a large value could artificially reduce the apparent error without any actual improvement in model performance. The pre-processing hyper-parameters we tuned included: (1) the additive constants $\epsilon_m$ and $\epsilon_n$; (2) the choice of an $l_1$ or $l_2$ norm for calculating total mass $m^{\text{tot}}$ (Equation~\ref{eq:09mtot}); (3) the scope of mass normalization for $u_1$ (across species, bins, or as a scalar); (4) the Box-Cox transformation exponents $\alpha_1$, $\alpha_2$, and $\alpha_3$; and (5) the application of zero-mean shift and unit-scaling components of the $\Z$-transform to $u_1$, $u_2$, and $u_3$, including their dimensional specificity and the use of a stabilization constant in the scaling denominator.

\begin{figure*}[t]
	\centering
	\includegraphics[width=0.98\linewidth]{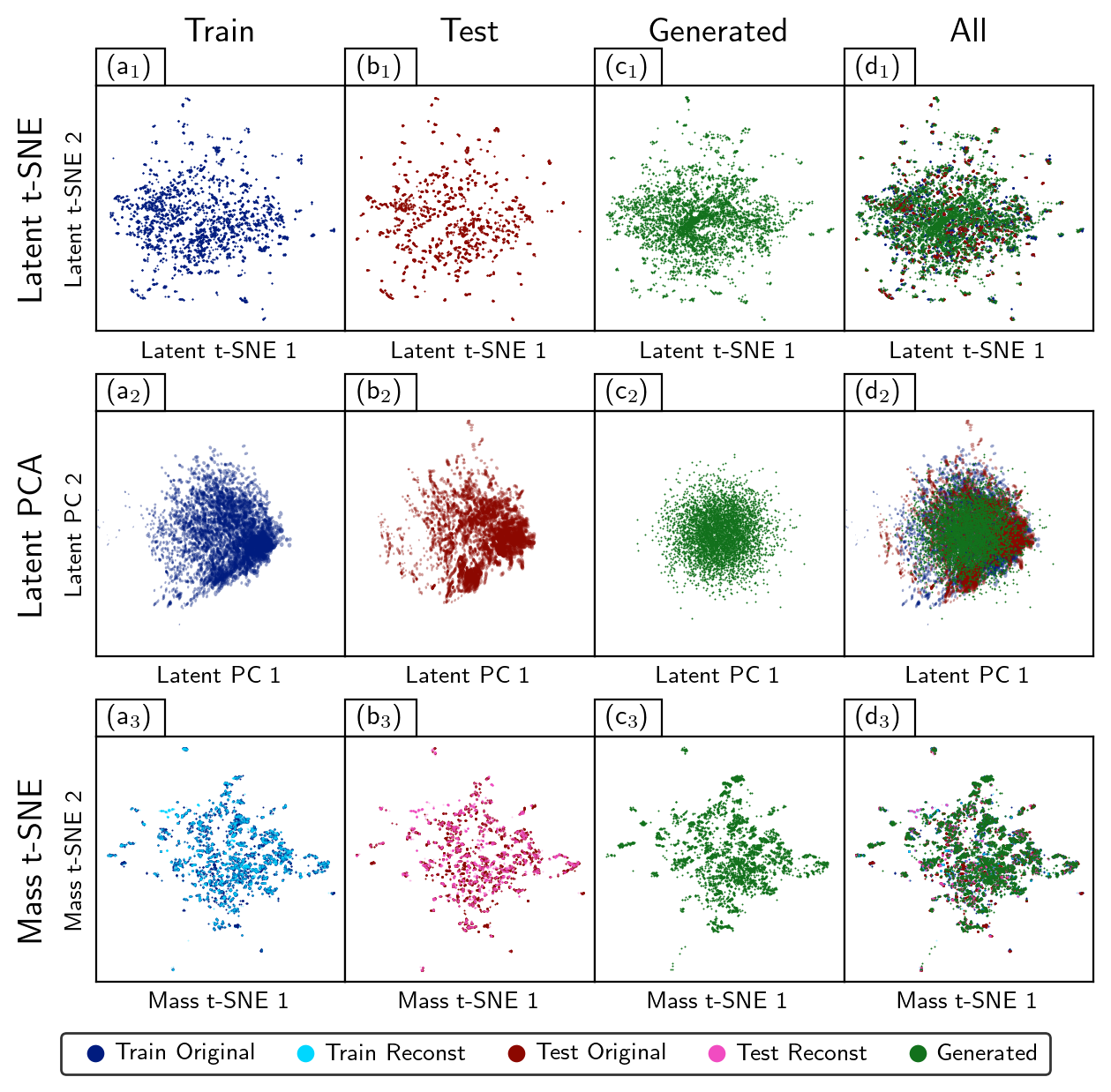}
	\caption{The multi-dimensional scaling of the samples in the latent and original data space. The \textit{top and middle rows} show the 2D t-SNE and PCA representation of the samples in the latent space. The \textit{bottom row} shows the 2D t-SNE representation of the speciated mass distributions in the original data space. The \textit{left, middle-left, and middle-right columns} are dedicated to the training, testing, and generated samples, respectively. The \textit{right column} is a compilation of all the training, testing, and generated samples in one plot. The horizontal and vertical axes are shared in each row.}\label{fig:10mdscnn}
\end{figure*}

After fixing the pre-processing parameters, we optimized the general model hyper-parameters. Starting from an initial guess, we performed a series of one-variable-at-a-time parameter sweeps, which are computationally tractable as they search for optimality along a single dimension at a time. Once a better hyper-parameter value was identified from a sweep, we manually updated it before proceeding to the next. This iterative process, akin to a manual batch coordinate descent, was repeated until no significant improvements were observed. This manual update strategy regularized the optimization, reduced the risk of over-optimization, and ensured that the final model retained a reasonable structural similarity to our initial configuration.

The realism metric (Equation~\ref{eq:06realism}) is highly sensitive to the choice of KL-divergence weights. Large $w_{\mu}$ weights can cause the model to over-regularize the squared L2-norm of the latent mean, $\|\mu\|_2^2$, forcing the encoded training data into a small region around the origin of the latent space. This concentration is not representative of the prior distribution, leaving large areas of the latent space devoid of training data and leading to unpredictable decoded samples. This effect is visible in Figure~\ref{fig:06hpstudy}(c$_2$), where large $w_{\mu}$ weights correspond to a significant increase in the realism error. Conversely, small $w_{\mu}$ weights can lead to under-regularization, causing the model to distribute the encoded data over a region much larger than the intended prior domain. While sampling from the central region of the latent space might still produce realistic samples, much of the original training data may be mapped to areas that are poorly represented by the prior. Such a distribution shift can cause the realism error to increase, since the metric considers the similarity of the entire training data population to the generated samples. Similar training dynamics are observed when setting the $w_{\sigma}$ weights too low or too high.

Figure~\ref{fig:06hpstudy} illustrates the effect of three key modeling
hyper-parameters. Overall, the CNN (conventional neural network) model
outperforms the MLP (multi-layer perceptron) in all metrics. As the latent
dimension increases, the CNN architecture shows an elbow-like improvement, while MLP's improvements are more linear. For both architectures, CCN
spectrum and optical property errors decrease as dimensionality increases,
without a clear plateau. We ultimately chose a 10-dimensional latent space to
achieve low errors in these specific diagnostics. For the $w_{\sigma}$ and
$w_{\mu}$ weights, we observed that higher weights tend to slightly worsen
per-sample reconstruction metrics. However, the population-based realism metric
exhibits a V-shaped pattern, indicating an optimal range for these weights. Our
results suggest that an incorrect $w_{\mu}$ weight is more detrimental to model
performance than an incorrect $w_{\sigma}$ weight.

\begin{figure*}[!thbp]
	\centering
	\includegraphics[page=1,width=0.98\linewidth]{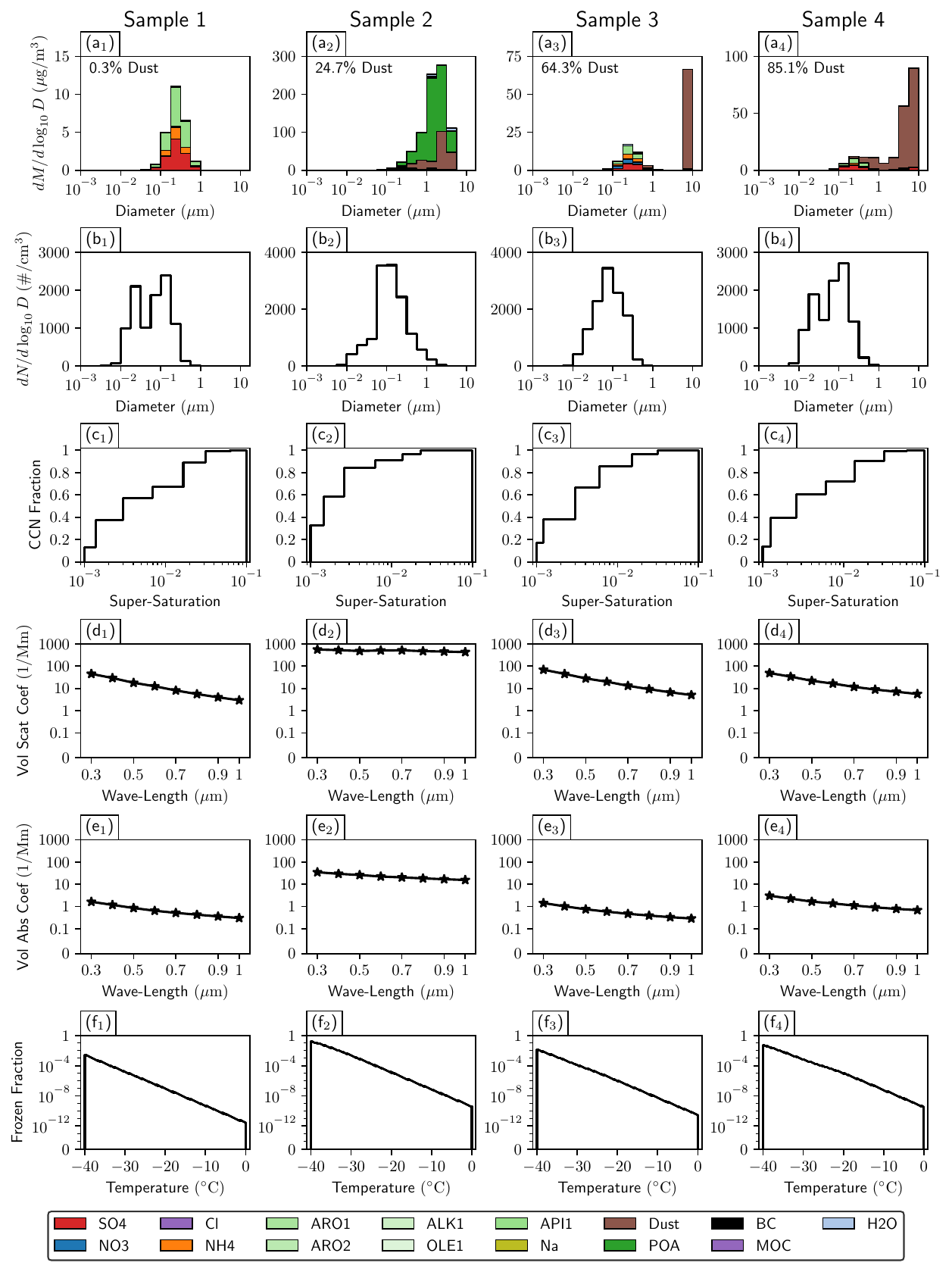}
	\caption{The aerosol diagnostics of four generated samples with different proportions of OIN mineral dust mass fraction. \textit{Each column} denotes a specific sample, and the rows correspond to different aerosol diagnostic variables. The samples are sorted based upon their OIN mass fraction from left to right. More examples are provided in Figures~\ref{fig:a07gendiag} and~\ref{fig:a08gendiag} of the appendix.}\label{fig:11gendiag}
\end{figure*}

Figure~\ref{fig:a01realism} in the appendix further explores the relationship
between realism, KL weights, and latent dimensionality under different realism
metric exponents, $\alpha$. Higher latent dimensions can yield better realism,
but this benefit is primarily observed with $\alpha=-1$ and is most significant
when moving from 2D to 3D. For $\alpha \ge 0$, lower-dimensional models perform
as well as higher-dimensional ones, a phenomenon attributable to PCA-like
behavior where principal components dominate the metric. Choosing a high latent
dimension can be risky without well-tuned KL weights; if hyper-parameters are
uncertain, a 2D latent space is often a safer choice. Because we were confident
in our hyper-parameters, we were able to use a higher latent dimensionality. As
shown in Figure~\ref{fig:a02realism}, which plots realism against latent
dimensionality, correctly tuned KL weights result in a clear downward trend in
realism error as dimensionality increases.

\begin{figure*}[t]
	\centering
	\includegraphics[width=0.98\linewidth]{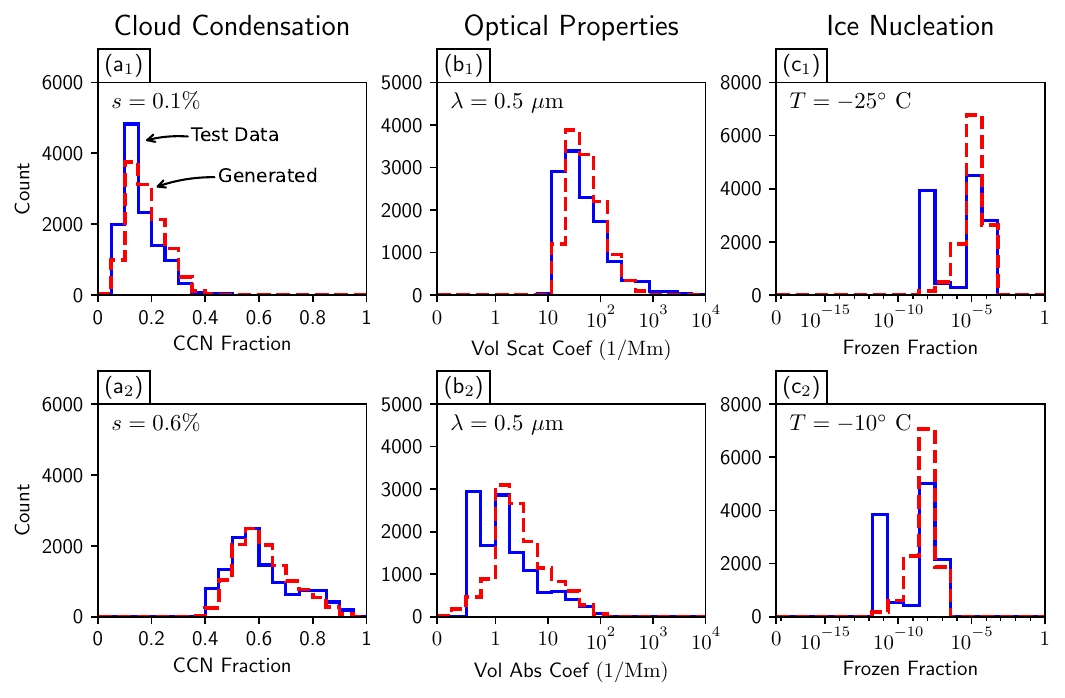}
	\caption{The generated aerosol diagnostics calibration plots. The blue and red histograms correspond to the held-out test (ground truth) and generated data, respectively. (a$_{1-2}$) show the CCN fraction histograms at $s=0.1\%$ and $s=0.6\%$ supersaturation levels, respectively. 
	(b$_{1-2}$) show the the volume scattering and absorption coefficient histograms at $\lambda=0.5 \, {\rm \mu m}$ wavelength.
	(c$_{1-2}$) show the frozen fraction histograms at $T=-25 \, {\rm ^{\circ} C}$ and $T=-10 \, {\rm ^{\circ} C}$, respectively.}\vspace{-7mm}\label{fig:12gencal}
\end{figure*}

\section{Results}

This section presents the results of our generative modeling framework. We first evaluate the model's ability to reconstruct aerosol states by examining both individual examples and collective error metrics across the test dataset. We then analyze the structure of the learned latent space to assess how the model organizes the aerosol data. Finally, we showcase the model's generative capabilities by presenting and analyzing newly generated aerosol samples.

\subsection{Aerosol Diagnostic Reconstruction Examples}

Figure~\ref{fig:08anecdiag} presents an anecdotal example comparing original and
reconstructed aerosol diagnostics; additional examples are provided in
Figures~\ref{fig:a05anecdiag} and~\ref{fig:a06anecdiag} in the appendix. While
the speciated mass relative error is quite small, the frozen fraction is
overestimated. This discrepancy is due to excess dust in the reconstructed
sample, which is small in absolute value (see i$_{11}$; both values are near
zero) but large in relative error (see j$_{11}$). Because the frozen fraction is
sensitive to dust, this large relative error in dust results in significant
frozen fraction error. However, because it is a small absolute error, this has
not been penalized by the model training procedure, which weights all species
equally. It is important to recall that the pre-processing was not tuned to
optimize any particular aerosol diagnostic, but rather to reconstruct all
species equally. Despite the error in frozen fraction, other diagnostics are
tracked accurately. To improve the frozen fraction accuracy, we could train the
model with a higher weight on the dust reconstruction error.

\subsection{Collective Aerosol Diagnostic Summaries}

Figure~\ref{fig:09smrytest} provides collective summary plots of the reconstruction performance on the test dataset; the corresponding plots for the training data are qualitatively similar, with slightly lower errors. Table~\ref{tab:rcnsterrs} details the reconstruction errors of the various aerosol diagnostics, as well. Among the various diagnostics, the CCN fractions are reconstructed with the highest accuracy, which is attributable to the binned representation of the aerosol data; the minor features visible above and below the diagonal indicate small lags or leads in the reconstructed CCN spectrum. The robust reconstruction of CCN spectra indicates that compressed aerosol representations could already be useful for studies of aerosol–cloud interactions, where droplet activation dominates the climate-relevant response of particle populations.

The volume scattering coefficient is reconstructed more easily than the absorption coefficient, as all species contribute to scattering, while absorption is entirely driven by black carbon. The moderate reconstruction skill for scattering and absorption coefficients highlights both the promise and the challenge of reduced-state representations: while average radiative effects are reasonably captured, composition-dependent absorption features—particularly for black carbon cores—remain more difficult to encode.

The frozen fraction is the most challenging diagnostic, exhibiting a systematic overestimation. This stems from the model’s tendency to slightly overestimate dust in some samples, which—because of the strong nonlinearity of ice activation—leads to large biases in INP concentrations. These errors in dust distribution arise because the training objective weighted all species equally, giving no special emphasis to trace components like dust that disproportionately control immersion freezing. Targeted weighting of species or diagnostics during training could substantially improve performance, highlighting an important direction for future optimization.

\subsection{Multi-Dimensional Scaling Plots}

Figure~\ref{fig:10mdscnn} shows t-SNE and PCA representations of both the latent and original data spaces. Thanks to the proper choice of KL weights, the latent representations of the training, testing, and generated samples all occupy a similar region of the space. This alignment is reassuring, as it suggests that the generated samples are likely to be similar in nature to the real data from the train and test splits. However, it is important to note that any potential distribution shift can only be definitively measured at the decoder's output. The t-SNE plots of the speciated mass further support the hypothesis that the generated samples are reasonably close to the train and test populations.

\subsection{Generated Aerosol Representations}

Figure~\ref{fig:11gendiag} shows anecdotal examples of generated aerosol representations and their corresponding diagnostics, with further examples provided in Figures~\ref{fig:a07gendiag} and~\ref{fig:a08gendiag} in the appendix. A key feature of these generated examples is that most of their dust content is concentrated in the larger size bins. This characteristic supports the hypothesis that the generated samples are realistic, as dust particles are typically found in the coarse mode. The generated samples display a variety of realistic modal structures. For instance, Sample 1 exhibits distinct Aitken and accumulation modes, while Sample 2 shows a merged Aitken-accumulation mode, a feature often observed in atmospheric measurements. Other samples demonstrate further diversity: Sample 3 contains a single mode in the Aitken to accumulation range alongside a coarse mode, and Sample 4 displays all three modes (Aitken, accumulation, and coarse). In the samples with a coarse mode, it is composed primarily of dust. Figure~\ref{fig:12gencal} verifies that the generated aerosol diagnostics have similar distributions to those of the test data.

\renewcommand{\arraystretch}{1}
\newcommand{\ZR}{\phantom{0}}
\begin{table}[t]
	\centering	
	\begin{tabular}{p{0.45\textwidth}p{0.08\textwidth}p{0.25\textwidth}}
	\midrule
	Aerosol Diagnostic Metric & \ZR Mean & [95\% CI] \\ \midrule
	CCN Spectrum Relative Error & \ZR4.64\% & [\ZR4.49\%, \ZR4.77\%] \\
	Scattering Log-Rel Error & \ZR2.22\% & [\ZR2.15\%, \ZR2.29\%] \\
	Absorption Log-Rel Error & 26.8\ZR\% & [26.0\ZR\%, 27.6\ZR\%] \\
	Frozen Fraction Log-Rel Error & \ZR4.10\% & [\ZR3.83\%, \ZR4.37\%] \\
	Speciated Mass Relative Error & 16.4\ZR\% & [15.8\ZR\%, 17.0\ZR\%] \\
	Number Relative Error & \ZR3.32\% & [\ZR3.18\%, \ZR3.42\%] \\
	Total Mass Relative Error & 12.3\ZR\% & [11.8\ZR\%, 12.8\ZR\%] \\
	Species Bulk Mass Relative Error & 11.1\ZR\% & [10.6\ZR\%, 11.7\ZR\%] \\
	\hline
	\end{tabular}
	\caption{The average errors of the trained model on the held-out test set.}
	\vspace{-5mm}
	\label{tab:rcnsterrs}
\end{table}

\section{Conclusions}

In this work, we presented a comprehensive framework for learning compact and robust generative representations of high-dimensional aerosol states using a variational autoencoder. Our findings demonstrate that detailed aerosol size and composition distributions, comprising hundreds of variables, can be compressed into a latent space of ten or fewer dimensions while preserving the fidelity of key climate-relevant diagnostics. We established that the model reconstructs cloud condensation nuclei spectra with high accuracy, optical properties moderately well, and ice nucleation properties with the most difficulty, highlighting areas for future model refinement. The introduction of a noise-resilience-based pre-processing optimization strategy and a novel realism metric based on the sliced Wasserstein distance provides a robust methodology for developing and tuning such generative models. This approach not only offers a pathway to significantly reduce the memory and computational burdens in large-scale climate simulations but also provides a structured method for generating realistic aerosol populations for a wide range of atmospheric studies.

By systematically comparing diagnostics that represent distinct aerosol–climate interactions—warm cloud activation, direct radiative forcing, and ice nucleation—this study highlights where generative modeling can most immediately benefit climate applications. The reliable recovery of CCN spectra points to near-term potential for improved representations of aerosol–cloud interactions, while the more limited skill for ice nucleation underscores an area requiring further development. Importantly, these difficulties do not reflect a fundamental limitation of the generative framework. Rather, they arise because the model training placed equal weight on all aerosol species, so trace components critical for ice nucleation were not preferentially optimized. Incorporating diagnostic- or species-specific weighting schemes offers a clear path to improve reconstruction of ice processes.
In this way, the framework provides not only a compression tool but also a roadmap for prioritizing aerosol processes in next-generation surrogate modeling efforts.

Looking forward, latent representations learned in this way could serve not just as compressed descriptors of aerosol states, but as dynamic variables whose temporal evolution can be modeled directly. This would open the door to surrogate aerosol models that capture the essential complexity of aerosol microphysics at a fraction of the computational cost, providing a pathway toward next-generation climate models that are both efficient and physically faithful.

%
%

\section*{Open Research Section}

The underlying data for this study can be accessed at \url{https://doi.org/10.13012/B2IDB-2774261_V1}. The code used for the analysis is available at \url{https://github.com/ehsansaleh/partnn}.

\acknowledgments

This work used GPU resources at the Delta supercomputer of the National Center for Supercomputing Applications through Allocation CIS220111 from the Advanced Cyberinfrastructure Coordination Ecosystem: Services and Support (ACCESS) program~\cite{boerner2023access}, which is supported by National Science Foundation grants \#2138259, \#2138286, \#2138307, \#2137603, and \#2138296.

This work was supported by the U.S. Department of Energy, Office of Science, Office of Biological and Environmental Research under Award Number DE-SC0022130, and the Laboratory Directed Research and Development program at Sandia National Laboratories. Sandia National Laboratories is a multimission laboratory managed and operated by National Technology and Engineering Solutions of Sandia LLC, a wholly owned subsidiary of Honeywell International Inc. for the U.S. Department of Energy’s National Nuclear Security Administration contract DE-NA0003525.

This paper describes objective technical results and analysis. Any subjective views or opinions that might be expressed in the paper do not necessarily represent the views of the U.S. Department of Energy or the United States Government.


\newpage
\appendix
\section{Supplementary Material}

\subsection{Probabilistic and Mathematical Notations} 
\newcommand{\zz}{z}
We denote expectations with $\EE_{P(\zz)}[h(\zz)]:=\int_{\zz} h(\zz)P(\zz)\diff \zz$. Note that only the random variable in the subscript (i.e., $\zz$) is eliminated after the expectation. The set of samples $\{x_1, \cdots x_n\}$ is denoted with $\xpi$. $h_{\theta}(x)$ denotes the output of a neural network, parameterized by $\theta$, on the input $x$. These notations and operators are summarized in Table~\ref{tab:mathnotation}.

\renewcommand{\arraystretch}{1}
\begin{table}[h]
	\centering	
	\begin{tabular}{p{0.15\textwidth}p{0.75\textwidth}}
	    \midrule
		Notation & Description \\ \midrule
		$f_{\theta}(u)$ & The encoder network parameterized by $\theta$ \\
		$g_{\theta}(u)$ & The decoder network parameterized by $\theta$ \\
		$N$ & Number of samples \\
		$A$ & The number of chemical species \\
		$B$ & The number of diameter bins in the histograms \\
		$m$ & The speciated mass distribution as a $A\times B$ matrix \\
		$n$ & The particle number distribution as a $B$-element vector \\
		$x$ & The generic aerosol sample consisting of $m$ and $n$\\
		$\TT$ & the pre-processing transformation \\
		$m^{\text{tot}}_b$ & The total mass of all species in the diameter bin $b$ \\
		$p_{a,b}$ & The mass fraction of species $a$ in the diameter bin $b$ \\
		$\epsilon_m, \epsilon_n$ & Small additive tolerances for mass and number distributions \\
		$\mu$ & The variational latent mean variable for a sample \\
		$\Sigma$ & The variational latent covariance variables for a sample \\
		$z$ & The latent variable representation \\
		$\Z$ & The standardization operator applying a zero-mean and unit-scaling transformation inferred over the training data \\
		$u$ & The pre-processed sample consisting of $u_1$, $u_2$, and $u_3$\\
		$\sigma$ & Generic variable for standard deviations or singular values \\
		$s$ & The critical relative humidity super-saturation level \\
		$\lambda$ & The optical properties wave-length \\
		$\xpi$ & Generic sample set \\
		$\EE_{P(\zz)}[h(\zz)]$ & Expectation of $h(z)$ over $z\sim P(\cdot)$\\
		$\SW$ & The sliced Wasserstein distance \\
		$\gen{z}$ & The generative latent variable sample \\
		$\gen{x}$ & The generated aerosol sample \\
		$x^\tst$ & The held-out aerosol sample \\\hline
	\end{tabular}	
	\caption{The mathematical notations used throughout the paper.}
	\label{tab:mathnotation}
\end{table}

\subsection{Additional Results}

\begin{figure*}[t]
	\centering
	\includegraphics[width=0.98\linewidth]{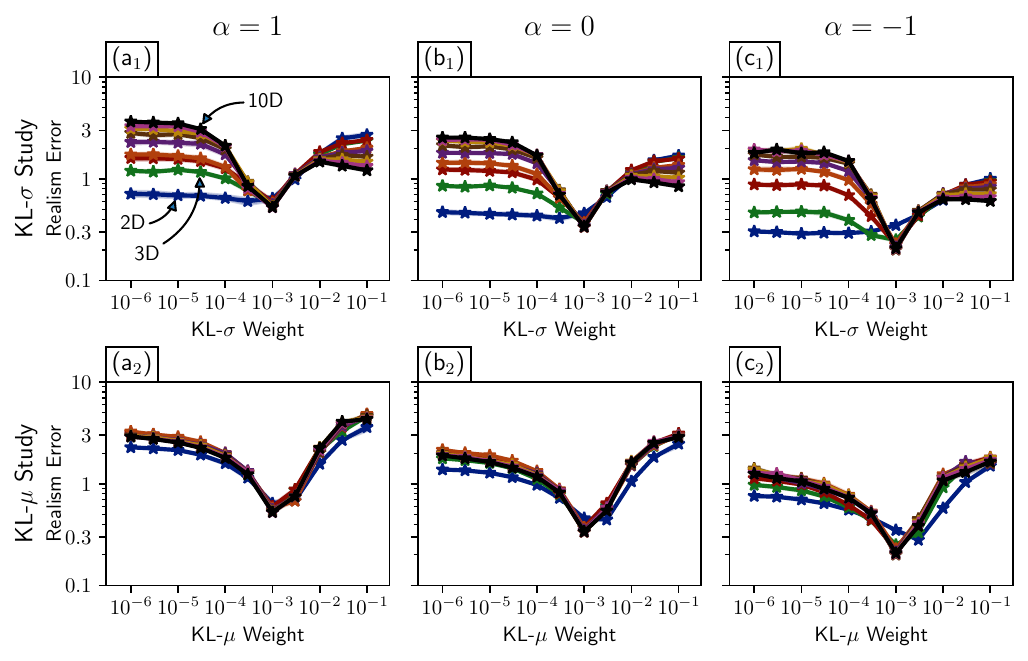}
	\caption{The realism metric vs the KL loss weights. The realism metric is defined as the sliced wasserstein distance between the held-out test data and the generated data by sampling latent variables from the $\mathcal{N}(0, I)$ distribution and passing them through the decoder. Each line denotes a specific latent dimensionality. The signular value exponent used for sampling the slicing directions is annotated in each plot.}\label{fig:a01realism}
\end{figure*}

Figure~\ref{fig:a01realism} illustrates the sensitivity of the model's generative realism to the Kullback-Leibler (KL) divergence loss weights. Each panel corresponds to a different weighting scheme ($\alpha$) for the sliced-Wasserstein distance calculation, which serves as the realism metric. Within each panel, the realism metric is plotted against the KL weights for models with varying latent space dimensionalities, represented by different colored lines. The distinct "V" shape of the curves demonstrates that there is an optimal range for the KL weights; values that are either too high or too low result in a poorer realism score, indicating less realistic generated samples.

\begin{figure*}[t]
	\centering
	\includegraphics[width=0.98\linewidth]{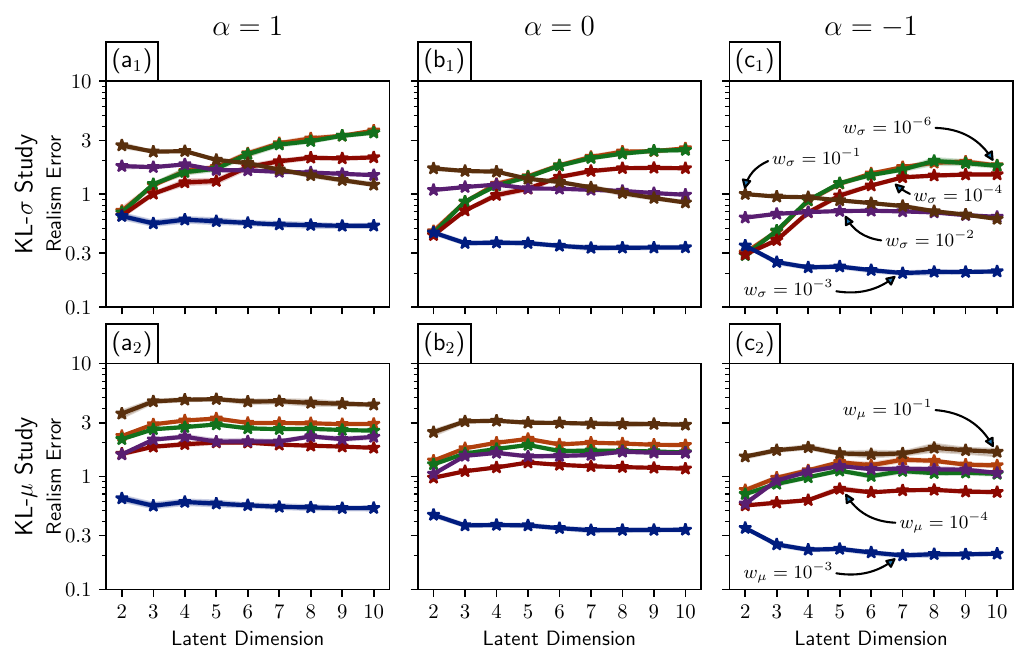}
	\caption{The realism metric vs the latent dimensionality. The realism metric is defined as the sliced wasserstein distance between the held-out test data and the generated data by sampling latent variables from the $\mathcal{N}(0, I)$ distribution and passing them through the decoder. Each line denotes a specific KL component weight. The signular value exponent used for sampling the slicing directions is annotated in each plot.}\label{fig:a02realism}
\end{figure*}

Figure~\ref{fig:a02realism} illustrates the relationship between the model's realism and its latent dimensionality. This figure essentially presents the same data as Figure~\ref{fig:a01realism} but with the axes swapped to highlight how realism changes as the number of latent dimensions increases. Each panel corresponds to a different weighting scheme ($\alpha$) for the sliced-Wasserstein distance. Within each panel, the different colored lines represent models trained with specific Kullback-Leibler (KL) divergence weights. The figure demonstrates that, for well-tuned KL weights, increasing the latent dimensionality generally improves the realism of the generated samples (i.e., the realism metric decreases). This trend is particularly evident when the realism metric is weighted to consider non-principal components of the data (e.g., $\alpha=-1$).

\begin{figure*}[t]
	\centering
	\includegraphics[width=0.98\linewidth]{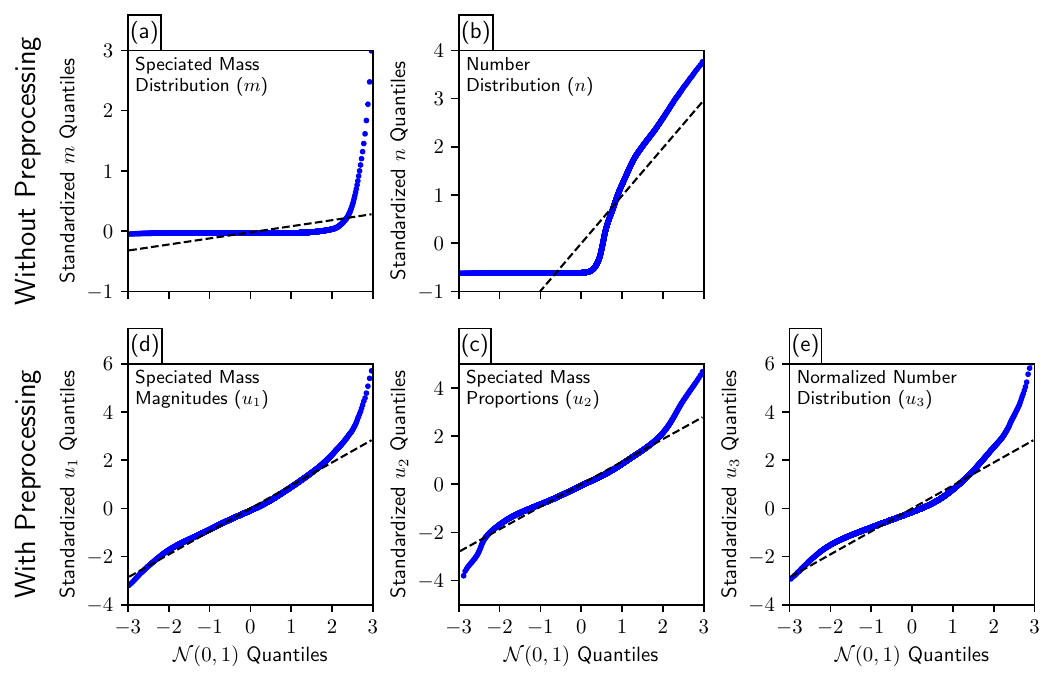}
	\caption{The Q-Q plots for the plain input values compared the pre-processed values using the optimal hyeper-parameters.}
	\label{fig:a03qqplot}
\end{figure*}

Figure~\ref{fig:a03qqplot} displays quantile-quantile (Q-Q) plots that compare the distributions of the simulated training with and without pre-processing against a standard normal distribution. The plots for the original data show significant deviation from the diagonal reference line, indicating that the data is highly non-Gaussian. In contrast, the plots for the pre-processed data align much more closely with the reference line, demonstrating that the transformation successfully makes the data distribution more normal. This normalization is a crucial step, as it helps the VAE model learn more effectively.

\begin{figure*}[t]
	\centering
	\includegraphics[page=2,width=0.98\linewidth]{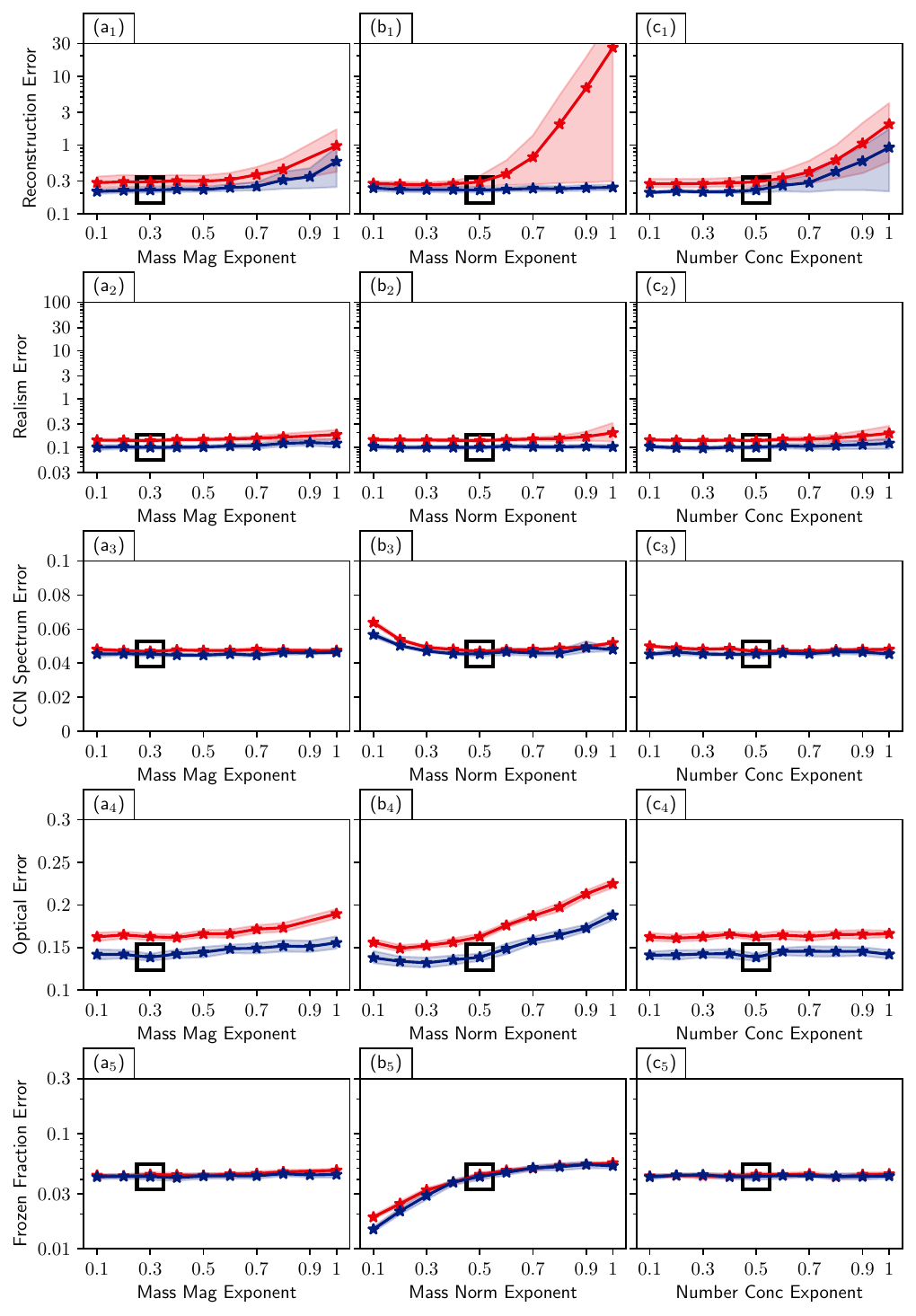}
	\vspace{-3mm}\caption{Ablating the effect of the mass and number concentration pre-processing exponents on various performance metrics.}\label{fig:a04hpstudy}
\end{figure*}

Figure~\ref{fig:a04hpstudy} illustrates the impact of key pre-processing hyper-parameters, specifically the Box-Cox exponents ($\alpha_1$, $\alpha_2$, and $\alpha_3$), on various model performance metrics. Each panel in the figure shows how a specific error metric—such as the reconstruction error for CCN, optical properties, or frozen fraction—changes as one of the exponents is varied while the others are held at their optimal values. The plots demonstrate that the model's performance is highly sensitive to these exponents. For example, the figure shows that using unit exponents (which corresponds to a near-linear transformation) results in higher errors, reinforcing the need for the non-linear pre-processing transformations and the optimization strategy used in this study. This figure justifies the simulation-based tuning approach by showing that a careful selection of these exponents is crucial for achieving optimal reconstruction accuracy.

\begin{figure*}[t]
	\centering
	\includegraphics[page=7,width=0.98\linewidth]{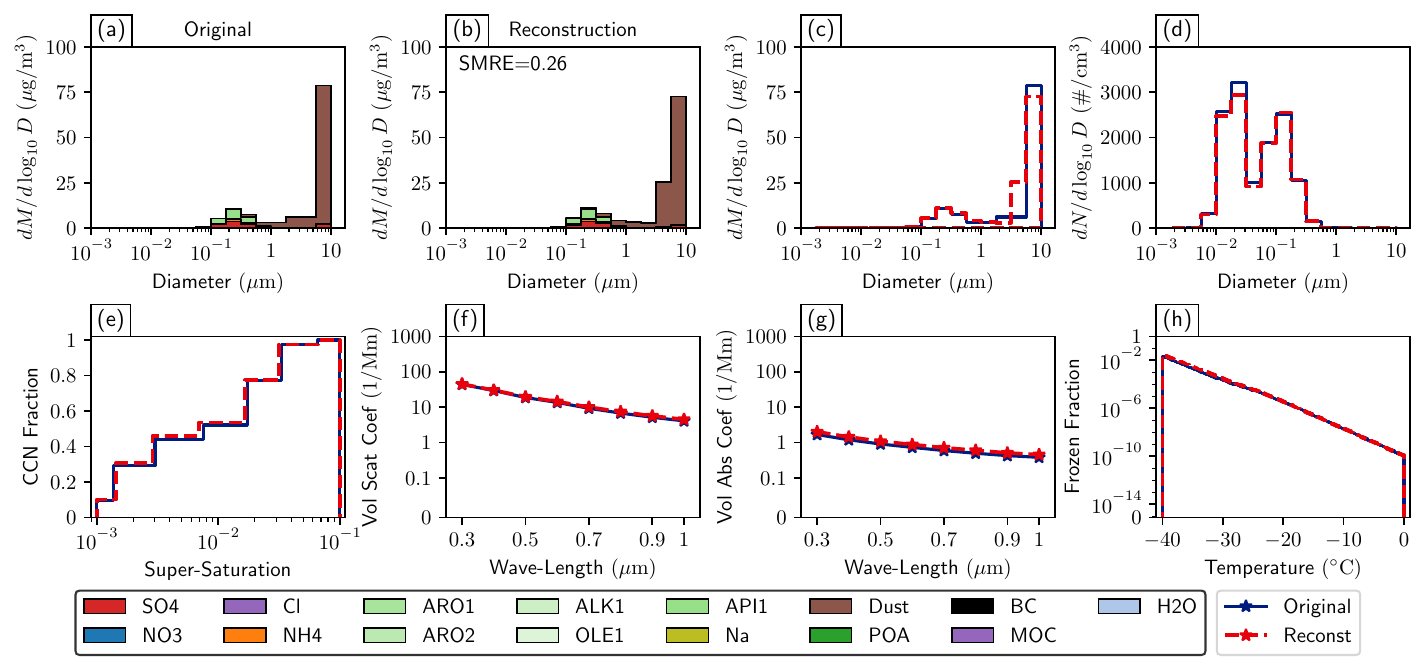}
	\includegraphics[page=8,width=0.98\linewidth]{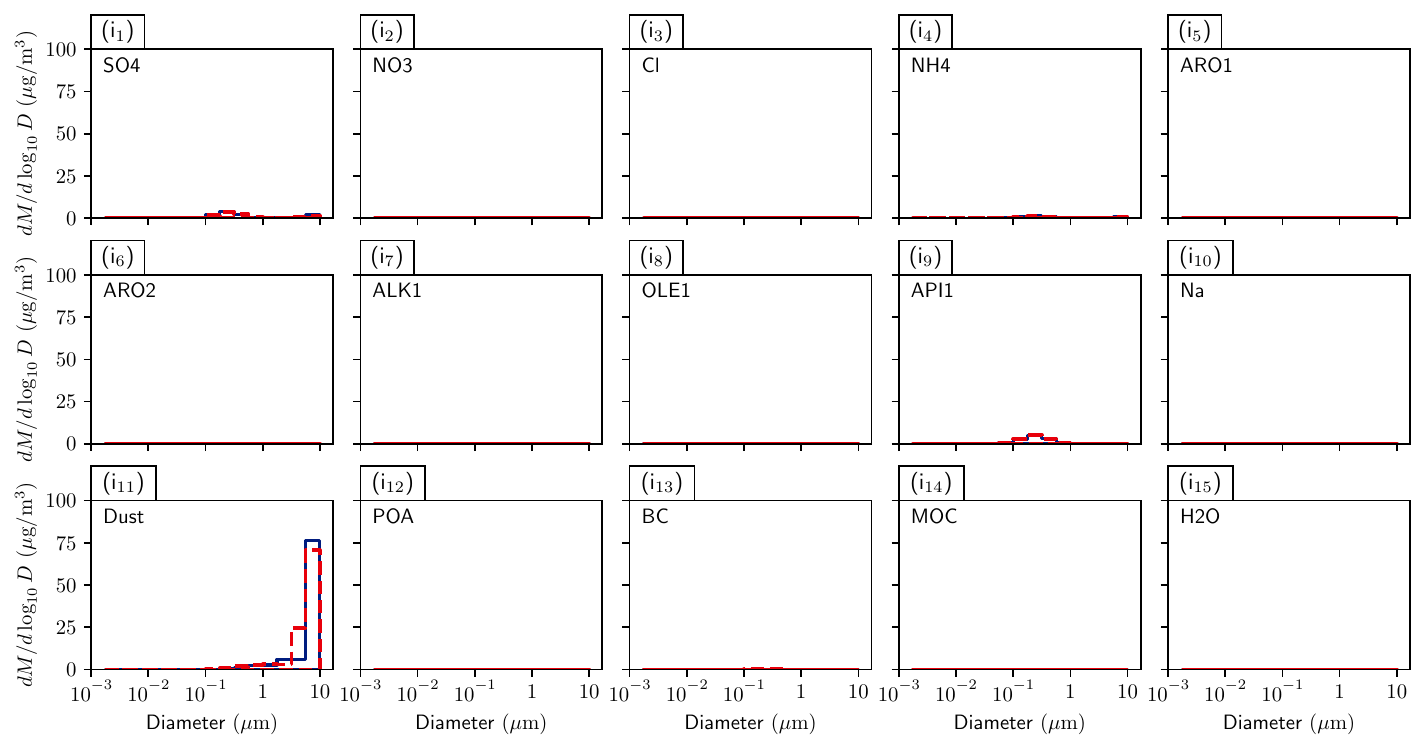}
	\includegraphics[page=9,width=0.98\linewidth]{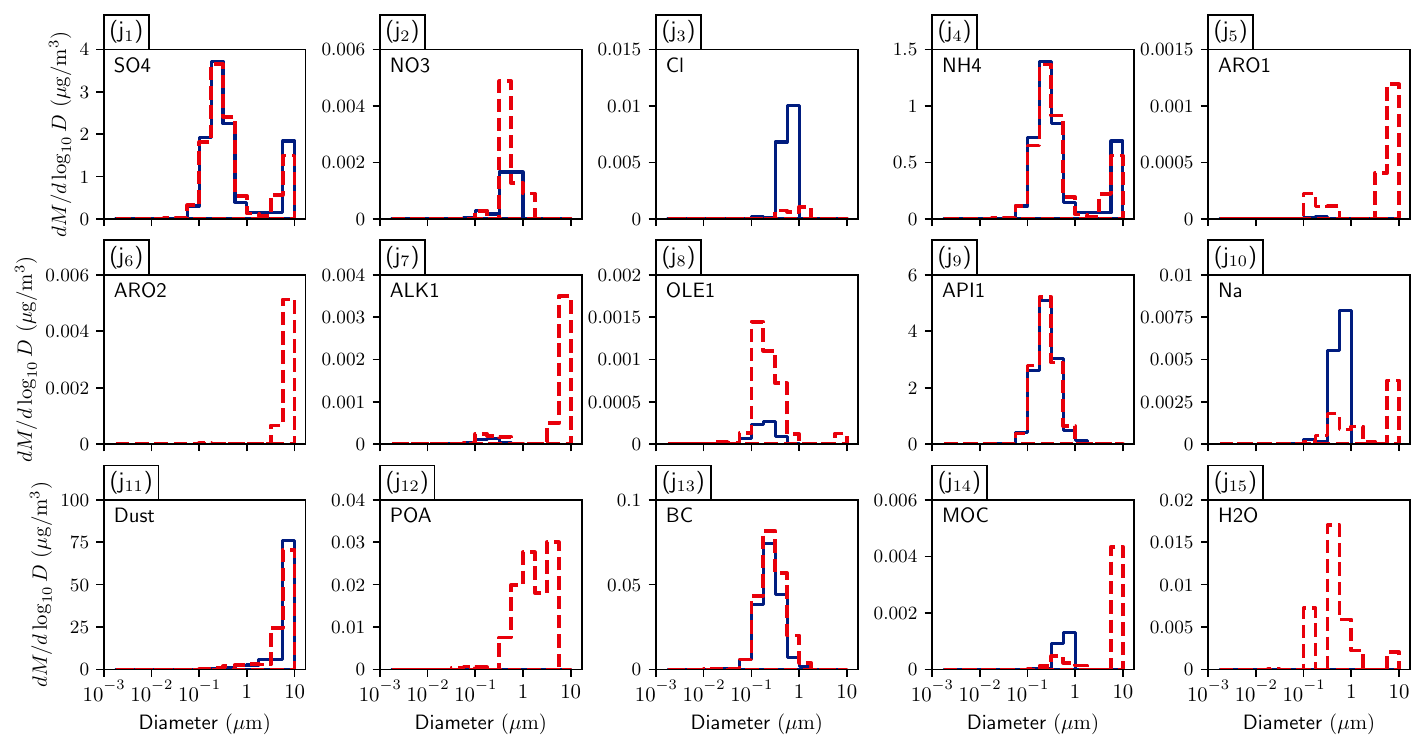}
	\caption{The aerosol diagnostics for another test sample with the same layout as Figure~\ref{fig:08anecdiag}. The particular sample in this figure has a speciated mass relative error of $0.26$.}\label{fig:a05anecdiag}
\end{figure*}

\begin{figure*}[t]
	\centering
	\includegraphics[page=13,width=0.98\linewidth]{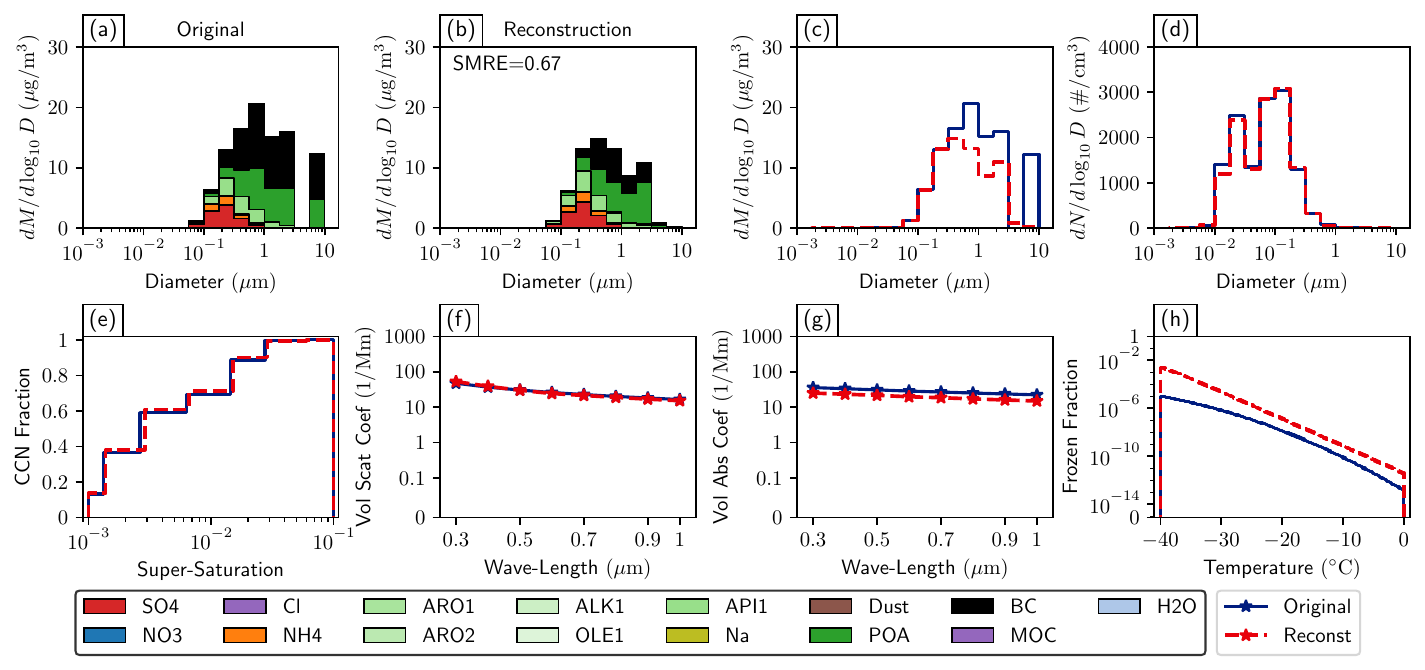}
	\includegraphics[page=14,width=0.98\linewidth]{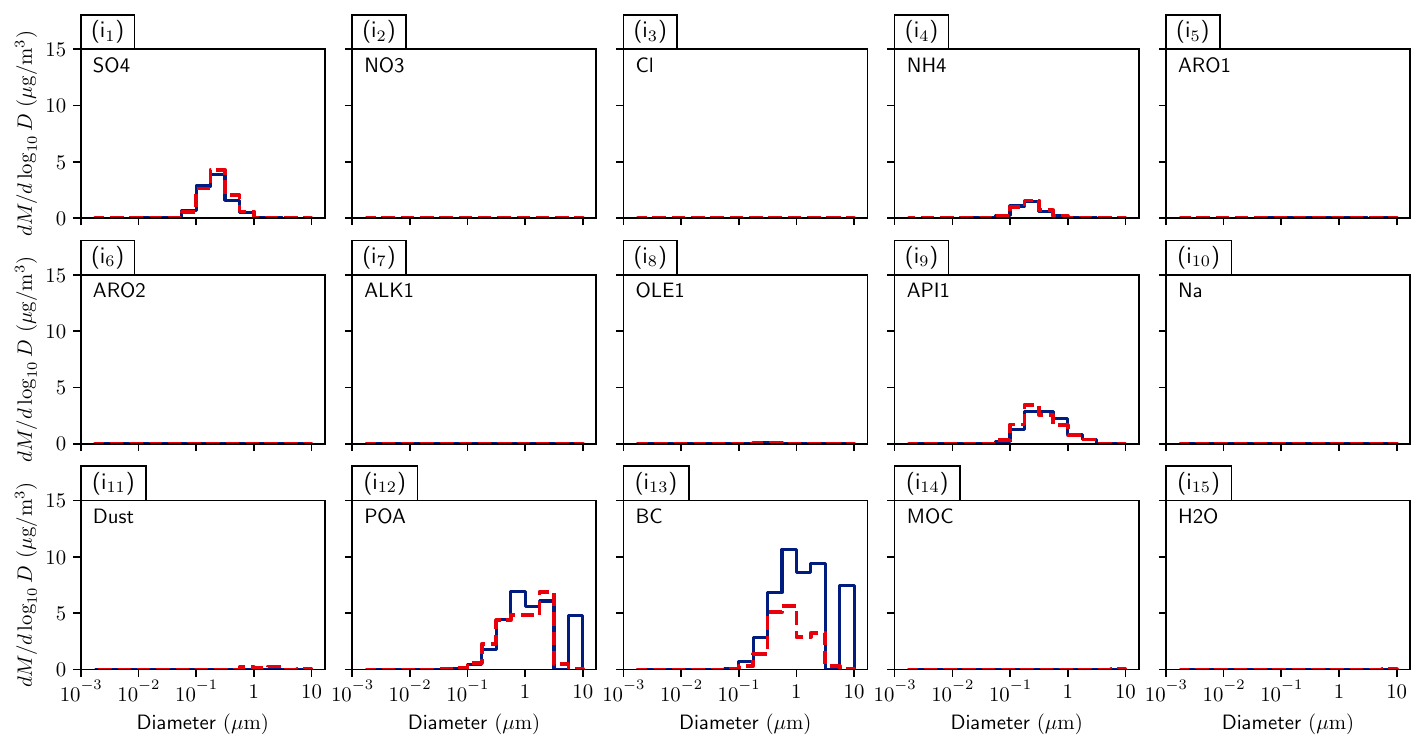}
	\includegraphics[page=15,width=0.98\linewidth]{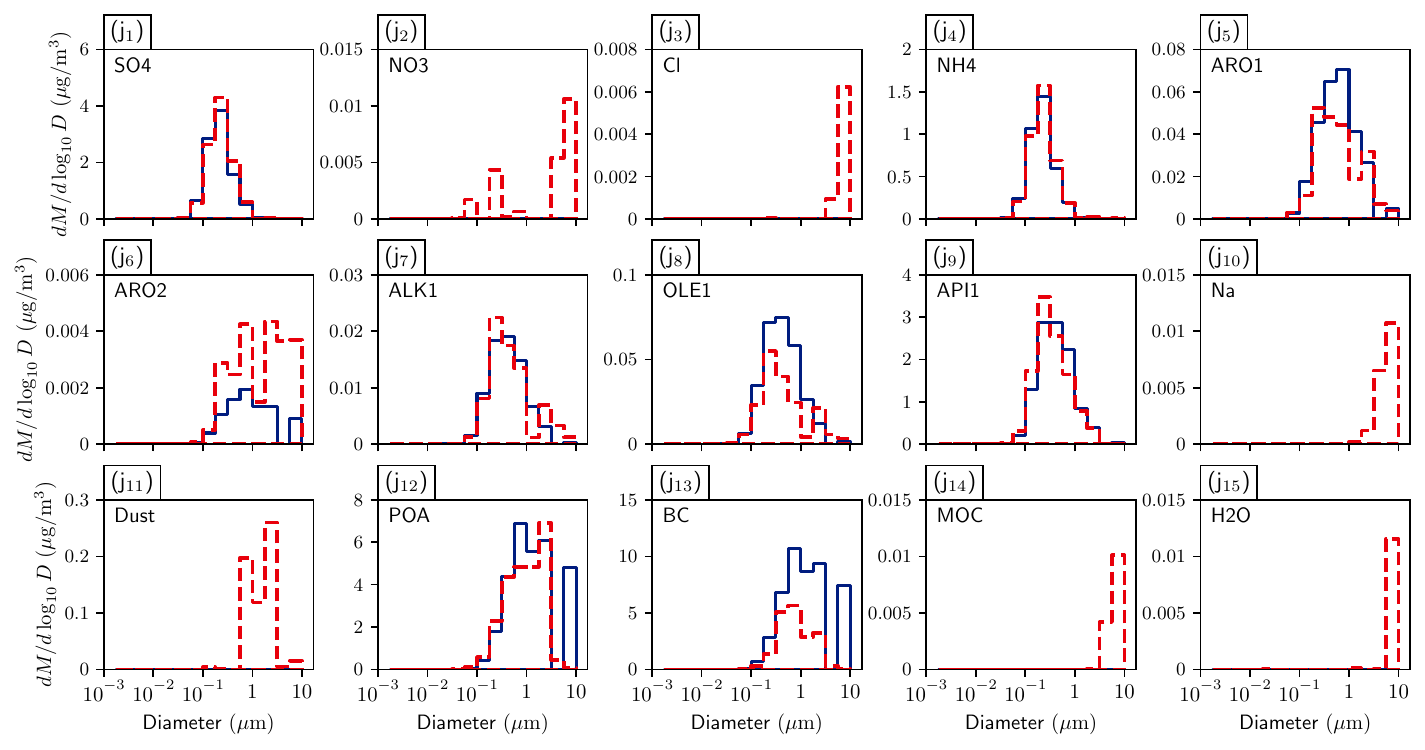}
	\caption{The aerosol diagnostics for another test sample with the same layout as Figure~\ref{fig:08anecdiag}. The particular sample in this figure has a speciated mass relative error of $0.67$.}\label{fig:a06anecdiag}
\end{figure*}

Figures~\ref{fig:a05anecdiag} and~\ref{fig:a06anecdiag} present other examples of the model's reconstruction performance on a single, specific test sample, following the same detailed layout as Figure~\ref{fig:08anecdiag}. This figure is intended to provide a more comprehensive view of the model's capabilities by showcasing its performance on a different case. These particular samples have a speciated mass relative error of 0.26 and 0.67, respectively, which is higher than the error for the sample in Figure~\ref{fig:08anecdiag}. This indicates that the figure illustrates a case where the reconstruction is less accurate, offering insight into the model's behavior on more challenging aerosol states.

\begin{figure*}[t]
	\centering
	\includegraphics[page=2,width=0.98\linewidth]{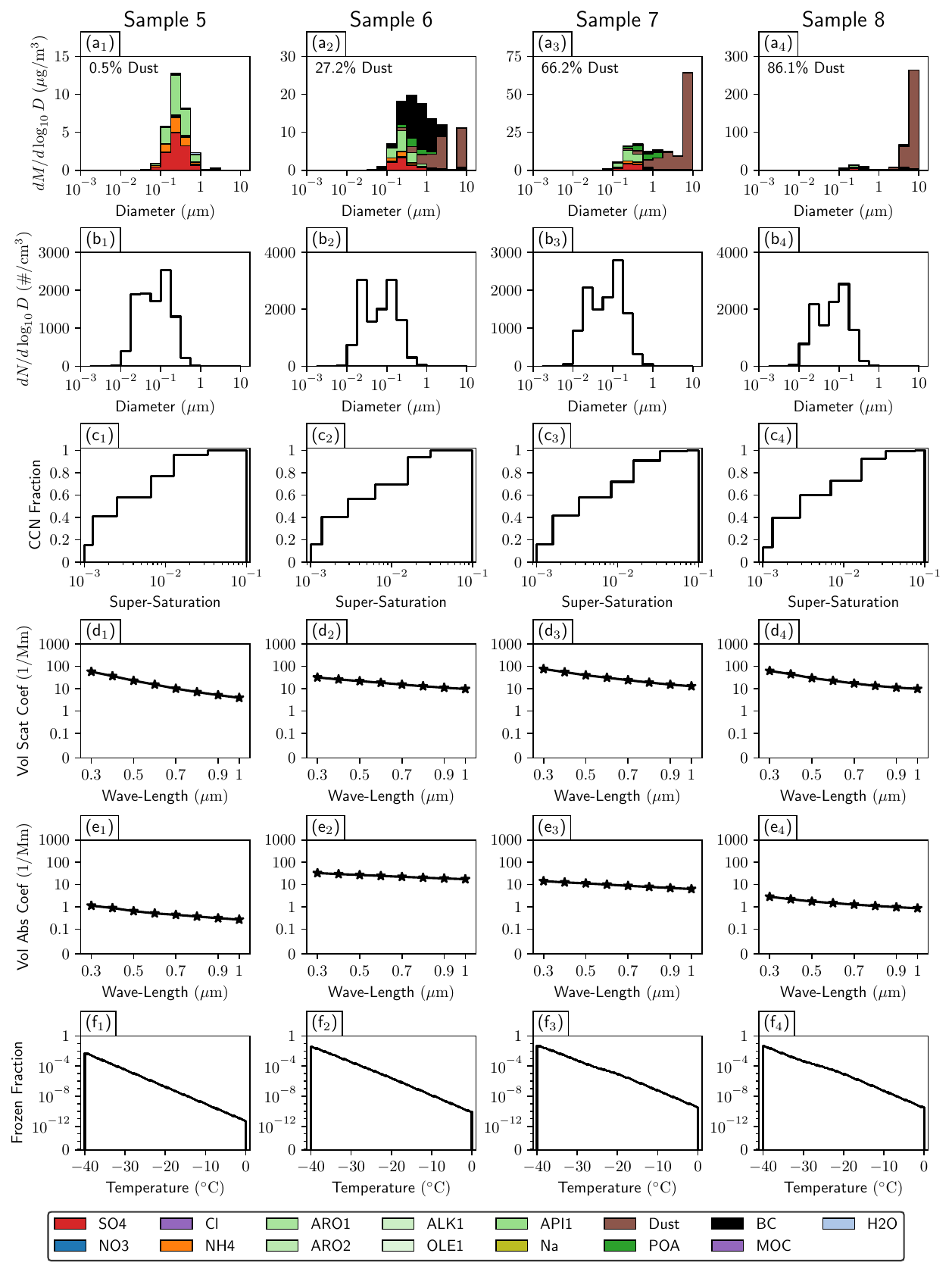}
	\caption{The aerosol diagnostics of another four generated samples with different proportions of mineral dust mass fraction, similar to Figure~\ref{fig:11gendiag}.}\label{fig:a07gendiag}
\end{figure*}

\begin{figure*}[t]
	\centering
	\includegraphics[page=3,width=0.98\linewidth]{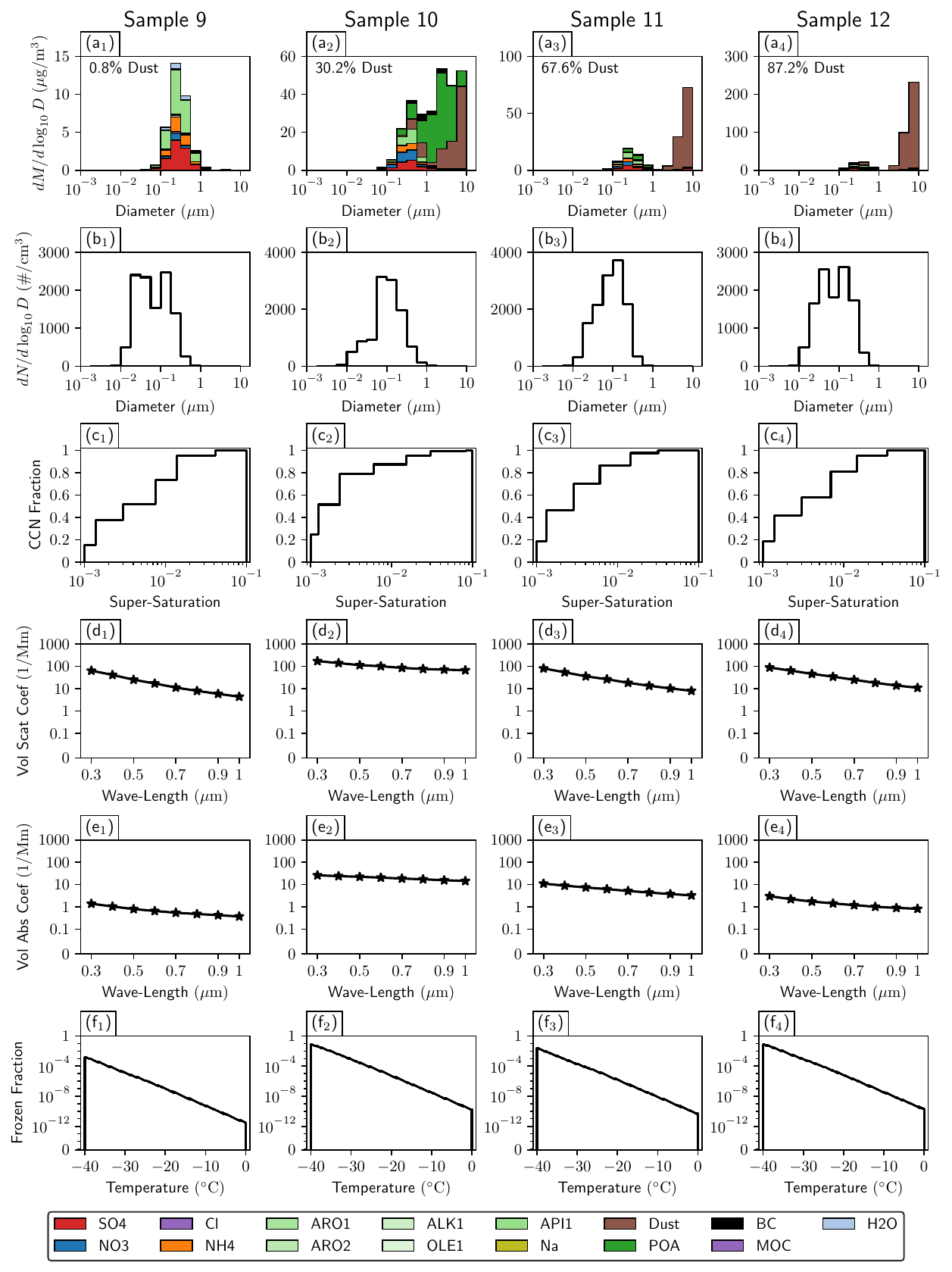}
	\caption{The aerosol diagnostics of another four generated samples with different proportions of mineral dust mass fraction, similar to Figure~\ref{fig:11gendiag}.}\label{fig:a08gendiag}
\end{figure*}

Figures~\ref{fig:a07gendiag} and~\ref{fig:a08gendiag} provide additional examples of generated aerosol populations to complement Figure~\ref{fig:11gendiag}. Each figure displays the full suite of aerosol diagnostics for 8 new, unique samples generated by the model. Following the same format as the main text figure, each column represents a distinct aerosol state, while the rows detail its properties, such as speciated mass distribution, number distribution, and various climate-relevant spectra. These additional examples further showcase the model's ability to generate a diverse range of physically plausible aerosol states with varied modal structures and chemical compositions, reinforcing the robustness of the generative framework.

\subsection{Implementation Details}

We used the Adam~\cite{kingma2014adam} optimizer with a learning rate of $10^{-3}$. The models were trained for 10000 iterations with a mini-batch size of 64. For the CNN models, we used a 4-layer convolutional model with 64 channels and a kernel size of 3, and the convolutional features were encoded into (and decoded from) the latent variables using a 2-layer MLP with 128 hidden units. We used the ReLU activation function for all neural models. For the optical properties calculations, we used 220 iterations in the Toon-Ackerman algorithm.

\begin{thebibliography}{}

\bibitem [\protect \citeauthoryear {%
Bauer%
\ \protect \BOthers {.}}{%
Bauer%
\ \protect \BOthers {.}}{%
{\protect \APACyear {2008}}%
}]{%
Bauer2008}
\APACinsertmetastar {%
Bauer2008}%
\begin{APACrefauthors}%
Bauer, S.%
, Wright, D.%
, Koch, D.%
, Lewis, E.%
, McGraw, R.%
, Chang, L\BHBI S.%
\BDBL {}Ruedy, R.%
\end{APACrefauthors}%
\unskip\
\newblock
\APACrefYearMonthDay{2008}{}{}.
\newblock
{\BBOQ}\APACrefatitle {{MATRIX} ({Multiconfiguration Aerosol TRacker of mIXing
  state}): an aerosol microphysical module for global atmospheric models}
  {{MATRIX} ({Multiconfiguration Aerosol TRacker of mIXing state}): an aerosol
  microphysical module for global atmospheric models}.{\BBCQ}
\newblock
\APACjournalVolNumPages{Atmos. Chem. Phys.}{8}{20}{6003--6035}.
\PrintBackRefs{\CurrentBib}

\bibitem [\protect \citeauthoryear {%
Binkowski%
\ \BBA {} Shankar%
}{%
Binkowski%
\ \BBA {} Shankar%
}{%
{\protect \APACyear {1995}}%
}]{%
Binkowski1995}
\APACinsertmetastar {%
Binkowski1995}%
\begin{APACrefauthors}%
Binkowski, F\BPBI S.%
\BCBT {}\ \BBA {} Shankar, U.%
\end{APACrefauthors}%
\unskip\
\newblock
\APACrefYearMonthDay{1995}{}{}.
\newblock
{\BBOQ}\APACrefatitle {The regional particulate matter model 1. {M}odel
  description and preliminary results} {The regional particulate matter model
  1. {M}odel description and preliminary results}.{\BBCQ}
\newblock
\APACjournalVolNumPages{J. Geophys. Res.}{100}{}{26191--26209}.
\PrintBackRefs{\CurrentBib}

\bibitem [\protect \citeauthoryear {%
Boerner%
, Deems%
, Furlani%
, Knuth%
\BCBL {}\ \BBA {} Towns%
}{%
Boerner%
\ \protect \BOthers {.}}{%
{\protect \APACyear {2023}}%
}]{%
boerner2023access}
\APACinsertmetastar {%
boerner2023access}%
\begin{APACrefauthors}%
Boerner, T\BPBI J.%
, Deems, S.%
, Furlani, T\BPBI R.%
, Knuth, S\BPBI L.%
\BCBL {}\ \BBA {} Towns, J.%
\end{APACrefauthors}%
\unskip\
\newblock
\APACrefYearMonthDay{2023}{}{}.
\newblock
{\BBOQ}\APACrefatitle {Access: Advancing innovation: {NSF}'s advanced
  cyberinfrastructure coordination ecosystem: Services and support} {Access:
  Advancing innovation: {NSF}'s advanced cyberinfrastructure coordination
  ecosystem: Services and support}.{\BBCQ}
\newblock
\BIn{} \APACrefbtitle {Practice and Experience in Advanced Research Computing}
  {Practice and experience in advanced research computing}\ (\BPGS\ 173--176).
\PrintBackRefs{\CurrentBib}

\bibitem [\protect \citeauthoryear {%
Ching%
, Fast%
, West%
\BCBL {}\ \BBA {} Riemer%
}{%
Ching%
\ \protect \BOthers {.}}{%
{\protect \APACyear {2017}}%
}]{%
Ching2017}
\APACinsertmetastar {%
Ching2017}%
\begin{APACrefauthors}%
Ching, J.%
, Fast, J.%
, West, M.%
\BCBL {}\ \BBA {} Riemer, N.%
\end{APACrefauthors}%
\unskip\
\newblock
\APACrefYearMonthDay{2017}{}{}.
\newblock
{\BBOQ}\APACrefatitle {Metrics to quantify the importance of mixing state for
  {CCN} activity} {Metrics to quantify the importance of mixing state for {CCN}
  activity}.{\BBCQ}
\newblock
\APACjournalVolNumPages{Atmos. Chem. Phys.}{17}{12}{7445}.
\PrintBackRefs{\CurrentBib}

\bibitem [\protect \citeauthoryear {%
Chung%
\ \BBA {} Seinfeld%
}{%
Chung%
\ \BBA {} Seinfeld%
}{%
{\protect \APACyear {2005}}%
}]{%
Chung2005}
\APACinsertmetastar {%
Chung2005}%
\begin{APACrefauthors}%
Chung, S\BPBI H.%
\BCBT {}\ \BBA {} Seinfeld, J\BPBI H.%
\end{APACrefauthors}%
\unskip\
\newblock
\APACrefYearMonthDay{2005}{}{}.
\newblock
{\BBOQ}\APACrefatitle {Climate response of direct radiative forcing of
  anthropogenic black carbon} {Climate response of direct radiative forcing of
  anthropogenic black carbon}.{\BBCQ}
\newblock
\APACjournalVolNumPages{J. Geophys. Res.}{110}{D11}{}.
\PrintBackRefs{\CurrentBib}

\bibitem [\protect \citeauthoryear {%
Ferracina%
\ \protect \BOthers {.}}{%
Ferracina%
\ \protect \BOthers {.}}{%
{\protect \APACyear {2025}}%
}]{%
ferracina2025learning}
\APACinsertmetastar {%
ferracina2025learning}%
\begin{APACrefauthors}%
Ferracina, F.%
, Beeler, P.%
, Halappanavar, M.%
, Krishnamoorthy, B.%
, Minutoli, M.%
\BCBL {}\ \BBA {} Fierce, L.%
\end{APACrefauthors}%
\unskip\
\newblock
\APACrefYearMonthDay{2025}{}{}.
\newblock
{\BBOQ}\APACrefatitle {Learning to Simulate Aerosol Dynamics with Graph Neural
  Networks} {Learning to simulate aerosol dynamics with graph neural
  networks}.{\BBCQ}
\newblock
\APACjournalVolNumPages{ACS ES\&T Air}{}{}{}.
\PrintBackRefs{\CurrentBib}

\bibitem [\protect \citeauthoryear {%
Fierce%
, Bond%
, Bauer%
, Mena%
\BCBL {}\ \BBA {} Riemer%
}{%
Fierce%
\ \protect \BOthers {.}}{%
{\protect \APACyear {2016}}%
}]{%
Fierce2016}
\APACinsertmetastar {%
Fierce2016}%
\begin{APACrefauthors}%
Fierce, L.%
, Bond, T\BPBI C.%
, Bauer, S\BPBI E.%
, Mena, F.%
\BCBL {}\ \BBA {} Riemer, N.%
\end{APACrefauthors}%
\unskip\
\newblock
\APACrefYearMonthDay{2016}{}{}.
\newblock
{\BBOQ}\APACrefatitle {Black carbon absorption at the global scale is affected
  by particle-scale diversity in composition} {Black carbon absorption at the
  global scale is affected by particle-scale diversity in composition}.{\BBCQ}
\newblock
\APACjournalVolNumPages{Nature communications}{7}{}{12361}.
\PrintBackRefs{\CurrentBib}

\bibitem [\protect \citeauthoryear {%
Gasparik%
\ \protect \BOthers {.}}{%
Gasparik%
\ \protect \BOthers {.}}{%
{\protect \APACyear {2020}}%
}]{%
Gasparik2020}
\APACinsertmetastar {%
Gasparik2020}%
\begin{APACrefauthors}%
Gasparik, J.%
, Ye, Q.%
, Curtis, J.%
, Presto, A.%
, Donahue, N.%
, Sullivan, R.%
\BDBL {}Riemer, N.%
\end{APACrefauthors}%
\unskip\
\newblock
\APACrefYearMonthDay{2020}{}{}.
\newblock
{\BBOQ}\APACrefatitle {Quantifying errors in the aerosol mixing-state index
  based on limited particle sample size} {Quantifying errors in the aerosol
  mixing-state index based on limited particle sample size}.{\BBCQ}
\newblock
\APACjournalVolNumPages{Aerosol Science and Technology}{54}{12}{1527--1541}.
\PrintBackRefs{\CurrentBib}

\bibitem [\protect \citeauthoryear {%
Goodfellow%
\ \protect \BOthers {.}}{%
Goodfellow%
\ \protect \BOthers {.}}{%
{\protect \APACyear {2014}}%
}]{%
goodfellow2014generative}
\APACinsertmetastar {%
goodfellow2014generative}%
\begin{APACrefauthors}%
Goodfellow, I\BPBI J.%
, Pouget-Abadie, J.%
, Mirza, M.%
, Xu, B.%
, Warde-Farley, D.%
, Ozair, S.%
\BDBL {}Bengio, Y.%
\end{APACrefauthors}%
\unskip\
\newblock
\APACrefYearMonthDay{2014}{}{}.
\newblock
{\BBOQ}\APACrefatitle {Generative adversarial nets} {Generative adversarial
  nets}.{\BBCQ}
\newblock
\APACjournalVolNumPages{Advances in neural information processing
  systems}{27}{}{}.
\PrintBackRefs{\CurrentBib}

\bibitem [\protect \citeauthoryear {%
Ho%
, Jain%
\BCBL {}\ \BBA {} Abbeel%
}{%
Ho%
\ \protect \BOthers {.}}{%
{\protect \APACyear {2020}}%
}]{%
ho2020denoising}
\APACinsertmetastar {%
ho2020denoising}%
\begin{APACrefauthors}%
Ho, J.%
, Jain, A.%
\BCBL {}\ \BBA {} Abbeel, P.%
\end{APACrefauthors}%
\unskip\
\newblock
\APACrefYearMonthDay{2020}{}{}.
\newblock
{\BBOQ}\APACrefatitle {Denoising diffusion probabilistic models} {Denoising
  diffusion probabilistic models}.{\BBCQ}
\newblock
\APACjournalVolNumPages{Advances in neural information processing
  systems}{33}{}{6840--6851}.
\PrintBackRefs{\CurrentBib}

\bibitem [\protect \citeauthoryear {%
Hoose%
\ \BBA {} M{\"o}hler%
}{%
Hoose%
\ \BBA {} M{\"o}hler%
}{%
{\protect \APACyear {2012}}%
}]{%
Hoose2012}
\APACinsertmetastar {%
Hoose2012}%
\begin{APACrefauthors}%
Hoose, C.%
\BCBT {}\ \BBA {} M{\"o}hler, O.%
\end{APACrefauthors}%
\unskip\
\newblock
\APACrefYearMonthDay{2012}{}{}.
\newblock
{\BBOQ}\APACrefatitle {Heterogeneous ice nucleation on atmospheric aerosols: A
  review of results from laboratory experiments} {Heterogeneous ice nucleation
  on atmospheric aerosols: A review of results from laboratory
  experiments}.{\BBCQ}
\newblock
\APACjournalVolNumPages{Atmospheric Chemistry and Physics}{12}{}{9817--9854}.
\newblock
\begin{APACrefDOI} \doi{10.5194/acp-12-9817-2012} \end{APACrefDOI}
\PrintBackRefs{\CurrentBib}

\bibitem [\protect \citeauthoryear {%
IPCC%
}{%
IPCC%
}{%
{\protect \APACyear {2021}}%
}]{%
IPCC2021}
\APACinsertmetastar {%
IPCC2021}%
\begin{APACrefauthors}%
IPCC.%
\end{APACrefauthors}%
\unskip\
\newblock
\APACrefYearMonthDay{2021}{}{}.
\newblock
{\BBOQ}\APACrefatitle {Climate Change 2021: The Physical Science Basis}
  {Climate change 2021: The physical science basis}.{\BBCQ}
\newblock
\BIn{} \APACrefbtitle {Contribution of Working Group {I} to the Sixth
  Assessment Report of the {Intergovernmental Panel on Climate Change}.}
  {Contribution of working group {I} to the sixth assessment report of the
  {Intergovernmental Panel on Climate Change}.}
\newblock
\APACaddressPublisher{}{Cambridge University Press}.
\newblock
\begin{APACrefURL} \url{https://www.ipcc.ch/report/ar6/wg1/} \end{APACrefURL}
\newblock
\begin{APACrefDOI} \doi{10.1017/9781009157896} \end{APACrefDOI}
\PrintBackRefs{\CurrentBib}

\bibitem [\protect \citeauthoryear {%
Jacobson%
}{%
Jacobson%
}{%
{\protect \APACyear {2001}}%
}]{%
Jacobson2001}
\APACinsertmetastar {%
Jacobson2001}%
\begin{APACrefauthors}%
Jacobson, M\BPBI Z.%
\end{APACrefauthors}%
\unskip\
\newblock
\APACrefYearMonthDay{2001}{}{}.
\newblock
{\BBOQ}\APACrefatitle {Strong radiative heating due to the mixing state of
  black carbon in atmospheric aerosols} {Strong radiative heating due to the
  mixing state of black carbon in atmospheric aerosols}.{\BBCQ}
\newblock
\APACjournalVolNumPages{Nature}{409}{6821}{695--697}.
\PrintBackRefs{\CurrentBib}

\bibitem [\protect \citeauthoryear {%
Jacobson%
}{%
Jacobson%
}{%
{\protect \APACyear {2005}}%
}]{%
Jacobson2005}
\APACinsertmetastar {%
Jacobson2005}%
\begin{APACrefauthors}%
Jacobson, M\BPBI Z.%
\end{APACrefauthors}%
\unskip\
\newblock
\APACrefYear{2005}.
\newblock
\APACrefbtitle {Fundamentals of Atmospheric Modeling} {Fundamentals of
  atmospheric modeling}\ (\PrintOrdinal{2nd}\ \BEd).
\newblock
\APACaddressPublisher{New York}{Cambridge University Press}.
\PrintBackRefs{\CurrentBib}

\bibitem [\protect \citeauthoryear {%
Kingma%
\ \BBA {} Ba%
}{%
Kingma%
\ \BBA {} Ba%
}{%
{\protect \APACyear {2014}}%
}]{%
kingma2014adam}
\APACinsertmetastar {%
kingma2014adam}%
\begin{APACrefauthors}%
Kingma, D\BPBI P.%
\BCBT {}\ \BBA {} Ba, J.%
\end{APACrefauthors}%
\unskip\
\newblock
\APACrefYearMonthDay{2014}{}{}.
\newblock
{\BBOQ}\APACrefatitle {Adam: A method for stochastic optimization} {Adam: A
  method for stochastic optimization}.{\BBCQ}
\newblock
\APACjournalVolNumPages{arXiv preprint arXiv:1412.6980}{}{}{}.
\PrintBackRefs{\CurrentBib}

\bibitem [\protect \citeauthoryear {%
Kingma%
\ \BBA {} Welling%
}{%
Kingma%
\ \BBA {} Welling%
}{%
{\protect \APACyear {2013}}%
}]{%
kingma2013auto}
\APACinsertmetastar {%
kingma2013auto}%
\begin{APACrefauthors}%
Kingma, D\BPBI P.%
\BCBT {}\ \BBA {} Welling, M.%
\end{APACrefauthors}%
\unskip\
\newblock
\APACrefYearMonthDay{2013}{}{}.
\newblock
{\BBOQ}\APACrefatitle {Auto-encoding variational bayes} {Auto-encoding
  variational bayes}.{\BBCQ}
\newblock
\APACjournalVolNumPages{arXiv preprint arXiv:1312.6114}{}{}{}.
\PrintBackRefs{\CurrentBib}

\bibitem [\protect \citeauthoryear {%
Kolouri%
, Nadjahi%
, {Ş}im{S}ekli%
, Badeau%
\BCBL {}\ \BBA {} Rohde%
}{%
Kolouri%
\ \protect \BOthers {.}}{%
{\protect \APACyear {2019}}%
}]{%
kolouri2019gsw}
\APACinsertmetastar {%
kolouri2019gsw}%
\begin{APACrefauthors}%
Kolouri, S.%
, Nadjahi, K.%
, {Ş}im{S}ekli, U.%
, Badeau, R.%
\BCBL {}\ \BBA {} Rohde, G\BPBI K.%
\end{APACrefauthors}%
\unskip\
\newblock
\APACrefYearMonthDay{2019}{}{}.
\newblock
{\BBOQ}\APACrefatitle {Generalized Sliced {W}asserstein Distances} {Generalized
  sliced {W}asserstein distances}.{\BBCQ}
\newblock
\BIn{} \APACrefbtitle {Advances in Neural Information Processing Systems}
  {Advances in neural information processing systems}\ (\BVOL~32).
\PrintBackRefs{\CurrentBib}

\bibitem [\protect \citeauthoryear {%
Lee%
\ \BBA {} Seung%
}{%
Lee%
\ \BBA {} Seung%
}{%
{\protect \APACyear {1999}}%
}]{%
lee1999learning}
\APACinsertmetastar {%
lee1999learning}%
\begin{APACrefauthors}%
Lee, D\BPBI D.%
\BCBT {}\ \BBA {} Seung, H\BPBI S.%
\end{APACrefauthors}%
\unskip\
\newblock
\APACrefYearMonthDay{1999}{}{}.
\newblock
{\BBOQ}\APACrefatitle {Learning the parts of objects by non-negative matrix
  factorization} {Learning the parts of objects by non-negative matrix
  factorization}.{\BBCQ}
\newblock
\APACjournalVolNumPages{nature}{401}{6755}{788--791}.
\PrintBackRefs{\CurrentBib}

\bibitem [\protect \citeauthoryear {%
Lipman%
, Chen%
, Ben-Hamu%
, Nickel%
\BCBL {}\ \BBA {} Le%
}{%
Lipman%
\ \protect \BOthers {.}}{%
{\protect \APACyear {2022}}%
}]{%
lipman2022flow}
\APACinsertmetastar {%
lipman2022flow}%
\begin{APACrefauthors}%
Lipman, Y.%
, Chen, R\BPBI T.%
, Ben-Hamu, H.%
, Nickel, M.%
\BCBL {}\ \BBA {} Le, M.%
\end{APACrefauthors}%
\unskip\
\newblock
\APACrefYearMonthDay{2022}{}{}.
\newblock
{\BBOQ}\APACrefatitle {Flow matching for generative modeling} {Flow matching
  for generative modeling}.{\BBCQ}
\newblock
\APACjournalVolNumPages{arXiv preprint arXiv:2210.02747}{}{}{}.
\PrintBackRefs{\CurrentBib}

\bibitem [\protect \citeauthoryear {%
McFiggans%
\ \protect \BOthers {.}}{%
McFiggans%
\ \protect \BOthers {.}}{%
{\protect \APACyear {2006}}%
}]{%
Mcfiggans2006}
\APACinsertmetastar {%
Mcfiggans2006}%
\begin{APACrefauthors}%
McFiggans, G.%
, Artaxo, P.%
, Baltensperger, U.%
, Coe, H.%
, Facchini, M\BPBI C.%
, Feingold, G.%
\BDBL {}others%
\end{APACrefauthors}%
\unskip\
\newblock
\APACrefYearMonthDay{2006}{}{}.
\newblock
{\BBOQ}\APACrefatitle {The effect of physical and chemical aerosol properties
  on warm cloud droplet activation} {The effect of physical and chemical
  aerosol properties on warm cloud droplet activation}.{\BBCQ}
\newblock
\APACjournalVolNumPages{Atmospheric Chemistry and Physics}{6}{9}{2593--2649}.
\PrintBackRefs{\CurrentBib}

\bibitem [\protect \citeauthoryear {%
Niemand%
\ \protect \BOthers {.}}{%
Niemand%
\ \protect \BOthers {.}}{%
{\protect \APACyear {2012}}%
}]{%
niemand2012particle}
\APACinsertmetastar {%
niemand2012particle}%
\begin{APACrefauthors}%
Niemand, M.%
, M{\"o}hler, O.%
, Vogel, B.%
, Vogel, H.%
, Hoose, C.%
, Connolly, P.%
\BDBL {}others%
\end{APACrefauthors}%
\unskip\
\newblock
\APACrefYearMonthDay{2012}{}{}.
\newblock
{\BBOQ}\APACrefatitle {A particle-surface-area-based parameterization of
  immersion freezing on desert dust particles} {A particle-surface-area-based
  parameterization of immersion freezing on desert dust particles}.{\BBCQ}
\newblock
\APACjournalVolNumPages{Journal of the Atmospheric
  Sciences}{69}{10}{3077--3092}.
\PrintBackRefs{\CurrentBib}

\bibitem [\protect \citeauthoryear {%
Pearson%
}{%
Pearson%
}{%
{\protect \APACyear {1901}}%
}]{%
pearson1901liii}
\APACinsertmetastar {%
pearson1901liii}%
\begin{APACrefauthors}%
Pearson, K.%
\end{APACrefauthors}%
\unskip\
\newblock
\APACrefYearMonthDay{1901}{}{}.
\newblock
{\BBOQ}\APACrefatitle {{LIII.} On lines and planes of closest fit to systems of
  points in space} {{LIII.} on lines and planes of closest fit to systems of
  points in space}.{\BBCQ}
\newblock
\APACjournalVolNumPages{The London, Edinburgh, and Dublin philosophical
  magazine and journal of science}{2}{11}{559--572}.
\PrintBackRefs{\CurrentBib}

\bibitem [\protect \citeauthoryear {%
P{\"o}schl%
}{%
P{\"o}schl%
}{%
{\protect \APACyear {2005}}%
}]{%
Poeschl2005}
\APACinsertmetastar {%
Poeschl2005}%
\begin{APACrefauthors}%
P{\"o}schl, U.%
\end{APACrefauthors}%
\unskip\
\newblock
\APACrefYearMonthDay{2005}{}{}.
\newblock
{\BBOQ}\APACrefatitle {Atmospheric aerosols: Composition, transformation,
  climate and health effects} {Atmospheric aerosols: Composition,
  transformation, climate and health effects}.{\BBCQ}
\newblock
\APACjournalVolNumPages{Angewandte Chemie International
  Edition}{44}{46}{7520--7540}.
\newblock
\begin{APACrefDOI} \doi{10.1002/anie.200501122} \end{APACrefDOI}
\PrintBackRefs{\CurrentBib}

\bibitem [\protect \citeauthoryear {%
Rezende%
\ \BBA {} Mohamed%
}{%
Rezende%
\ \BBA {} Mohamed%
}{%
{\protect \APACyear {2015}}%
}]{%
rezende2015variational}
\APACinsertmetastar {%
rezende2015variational}%
\begin{APACrefauthors}%
Rezende, D.%
\BCBT {}\ \BBA {} Mohamed, S.%
\end{APACrefauthors}%
\unskip\
\newblock
\APACrefYearMonthDay{2015}{}{}.
\newblock
{\BBOQ}\APACrefatitle {Variational inference with normalizing flows}
  {Variational inference with normalizing flows}.{\BBCQ}
\newblock
\BIn{} \APACrefbtitle {International conference on machine learning}
  {International conference on machine learning}\ (\BPGS\ 1530--1538).
\PrintBackRefs{\CurrentBib}

\bibitem [\protect \citeauthoryear {%
Riemer%
, Ault%
, West%
, Craig%
\BCBL {}\ \BBA {} Curtis%
}{%
Riemer%
\ \protect \BOthers {.}}{%
{\protect \APACyear {2019}}%
}]{%
Riemer2019}
\APACinsertmetastar {%
Riemer2019}%
\begin{APACrefauthors}%
Riemer, N.%
, Ault, A\BPBI P.%
, West, M.%
, Craig, R\BPBI L.%
\BCBL {}\ \BBA {} Curtis, J\BPBI H.%
\end{APACrefauthors}%
\unskip\
\newblock
\APACrefYearMonthDay{2019}{}{}.
\newblock
{\BBOQ}\APACrefatitle {Aerosol mixing state: Measurements, modeling, and
  impacts} {Aerosol mixing state: Measurements, modeling, and impacts}.{\BBCQ}
\newblock
\APACjournalVolNumPages{Reviews of Geophysics}{57}{2}{187--249}.
\newblock
\begin{APACrefDOI} \doi{10.1029/2018RG000615} \end{APACrefDOI}
\PrintBackRefs{\CurrentBib}

\bibitem [\protect \citeauthoryear {%
Riemer%
, West%
, Zaveri%
\BCBL {}\ \BBA {} Easter%
}{%
Riemer%
\ \protect \BOthers {.}}{%
{\protect \APACyear {2010}}%
}]{%
riemer2010}
\APACinsertmetastar {%
riemer2010}%
\begin{APACrefauthors}%
Riemer, N.%
, West, M.%
, Zaveri, R.%
\BCBL {}\ \BBA {} Easter, R.%
\end{APACrefauthors}%
\unskip\
\newblock
\APACrefYearMonthDay{2010}{}{}.
\newblock
{\BBOQ}\APACrefatitle {Estimating black carbon aging time-scales with a
  particle-resolved aerosol model} {Estimating black carbon aging time-scales
  with a particle-resolved aerosol model}.{\BBCQ}
\newblock
\APACjournalVolNumPages{Journal of Aerosol Science}{41}{1}{143--158}.
\PrintBackRefs{\CurrentBib}

\bibitem [\protect \citeauthoryear {%
Riemer%
, West%
, Zaveri%
\BCBL {}\ \BBA {} Easter%
}{%
Riemer%
\ \protect \BOthers {.}}{%
{\protect \APACyear {2009}}%
}]{%
Riemer2009}
\APACinsertmetastar {%
Riemer2009}%
\begin{APACrefauthors}%
Riemer, N.%
, West, M.%
, Zaveri, R\BPBI A.%
\BCBL {}\ \BBA {} Easter, R\BPBI C.%
\end{APACrefauthors}%
\unskip\
\newblock
\APACrefYearMonthDay{2009}{}{}.
\newblock
{\BBOQ}\APACrefatitle {Simulating the evolution of soot mixing state with a
  particle-resolved aerosol model} {Simulating the evolution of soot mixing
  state with a particle-resolved aerosol model}.{\BBCQ}
\newblock
\APACjournalVolNumPages{Journal of Geophysical Research:
  Atmospheres}{114}{D9}{}.
\PrintBackRefs{\CurrentBib}

\bibitem [\protect \citeauthoryear {%
Schill%
\ \protect \BOthers {.}}{%
Schill%
\ \protect \BOthers {.}}{%
{\protect \APACyear {2020}}%
}]{%
Schill2020}
\APACinsertmetastar {%
Schill2020}%
\begin{APACrefauthors}%
Schill, G\BPBI P.%
, DeMott, P\BPBI J.%
, Emerson, E\BPBI W.%
, Rauker, A\BPBI M\BPBI C.%
, Kodros, J\BPBI K.%
, Suski, K\BPBI J.%
\BDBL {}others%
\end{APACrefauthors}%
\unskip\
\newblock
\APACrefYearMonthDay{2020}{}{}.
\newblock
{\BBOQ}\APACrefatitle {The contribution of black carbon to global ice
  nucleating particle concentrations relevant to mixed-phase clouds} {The
  contribution of black carbon to global ice nucleating particle concentrations
  relevant to mixed-phase clouds}.{\BBCQ}
\newblock
\APACjournalVolNumPages{Proceedings of the National Academy of
  Sciences}{117}{37}{22705--22711}.
\PrintBackRefs{\CurrentBib}

\bibitem [\protect \citeauthoryear {%
Vignati%
, Wilson%
\BCBL {}\ \BBA {} Stier%
}{%
Vignati%
\ \protect \BOthers {.}}{%
{\protect \APACyear {2004}}%
}]{%
Vignati2004}
\APACinsertmetastar {%
Vignati2004}%
\begin{APACrefauthors}%
Vignati, E.%
, Wilson, J.%
\BCBL {}\ \BBA {} Stier, P.%
\end{APACrefauthors}%
\unskip\
\newblock
\APACrefYearMonthDay{2004}{}{}.
\newblock
{\BBOQ}\APACrefatitle {M7: An efficient size-resolved aerosol microphysics
  module for large-scale aerosol transport models} {M7: An efficient
  size-resolved aerosol microphysics module for large-scale aerosol transport
  models}.{\BBCQ}
\newblock
\APACjournalVolNumPages{J. Geophys. Res.}{109}{D22}{}.
\newblock
\begin{APACrefDOI} \doi{10.1029/2003JD004485} \end{APACrefDOI}
\PrintBackRefs{\CurrentBib}

\bibitem [\protect \citeauthoryear {%
Whitby%
\ \BBA {} McMurry%
}{%
Whitby%
\ \BBA {} McMurry%
}{%
{\protect \APACyear {1997}}%
}]{%
Whitby1997}
\APACinsertmetastar {%
Whitby1997}%
\begin{APACrefauthors}%
Whitby, E\BPBI R.%
\BCBT {}\ \BBA {} McMurry, P\BPBI H.%
\end{APACrefauthors}%
\unskip\
\newblock
\APACrefYearMonthDay{1997}{}{}.
\newblock
{\BBOQ}\APACrefatitle {Modal aerosol dynamics modeling} {Modal aerosol dynamics
  modeling}.{\BBCQ}
\newblock
\APACjournalVolNumPages{Aerosol Sci. Technol.}{27}{6}{673--688}.
\newblock
\begin{APACrefDOI} \doi{10.1080/02786829708965504} \end{APACrefDOI}
\PrintBackRefs{\CurrentBib}

\bibitem [\protect \citeauthoryear {%
Yao%
, Curtis%
, Ching%
, Zheng%
\BCBL {}\ \BBA {} Riemer%
}{%
Yao%
\ \protect \BOthers {.}}{%
{\protect \APACyear {2022}}%
}]{%
yao2022quantifying}
\APACinsertmetastar {%
yao2022quantifying}%
\begin{APACrefauthors}%
Yao, Y.%
, Curtis, J.%
, Ching, J.%
, Zheng, Z.%
\BCBL {}\ \BBA {} Riemer, N.%
\end{APACrefauthors}%
\unskip\
\newblock
\APACrefYearMonthDay{2022}{}{}.
\newblock
{\BBOQ}\APACrefatitle {Quantifying the effects of mixing state on aerosol
  optical properties} {Quantifying the effects of mixing state on aerosol
  optical properties}.{\BBCQ}
\newblock
\APACjournalVolNumPages{Atmospheric Chemistry and Physics}{2022}{}{9265-9282}.
\PrintBackRefs{\CurrentBib}

\bibitem [\protect \citeauthoryear {%
Zaveri%
, Barnard%
, Easter%
, Riemer%
\BCBL {}\ \BBA {} West%
}{%
Zaveri%
\ \protect \BOthers {.}}{%
{\protect \APACyear {2010}}%
}]{%
Zaveri2010}
\APACinsertmetastar {%
Zaveri2010}%
\begin{APACrefauthors}%
Zaveri, R\BPBI A.%
, Barnard, J\BPBI C.%
, Easter, R\BPBI C.%
, Riemer, N.%
\BCBL {}\ \BBA {} West, M.%
\end{APACrefauthors}%
\unskip\
\newblock
\APACrefYearMonthDay{2010}{}{}.
\newblock
{\BBOQ}\APACrefatitle {Particle-resolved simulation of aerosol size,
  composition, mixing state, and the associated optical and cloud condensation
  nuclei activation properties in an evolving urban plume} {Particle-resolved
  simulation of aerosol size, composition, mixing state, and the associated
  optical and cloud condensation nuclei activation properties in an evolving
  urban plume}.{\BBCQ}
\newblock
\APACjournalVolNumPages{J. Geophys. Res.}{115}{D17}{}.
\PrintBackRefs{\CurrentBib}

\bibitem [\protect \citeauthoryear {%
Zaveri%
, Easter%
, Fast%
\BCBL {}\ \BBA {} Peters%
}{%
Zaveri%
\ \protect \BOthers {.}}{%
{\protect \APACyear {2008}}%
}]{%
Zaveri2008}
\APACinsertmetastar {%
Zaveri2008}%
\begin{APACrefauthors}%
Zaveri, R\BPBI A.%
, Easter, R\BPBI C.%
, Fast, J\BPBI D.%
\BCBL {}\ \BBA {} Peters, L\BPBI K.%
\end{APACrefauthors}%
\unskip\
\newblock
\APACrefYearMonthDay{2008}{}{}.
\newblock
{\BBOQ}\APACrefatitle {Model for simulating aerosol interactions and chemistry
  ({MOSAIC})} {Model for simulating aerosol interactions and chemistry
  ({MOSAIC})}.{\BBCQ}
\newblock
\APACjournalVolNumPages{Journal of Geophysical Research:
  Atmospheres}{113}{D13}{}.
\newblock
\begin{APACrefDOI} \doi{10.1029/2007JD008782} \end{APACrefDOI}
\PrintBackRefs{\CurrentBib}

\bibitem [\protect \citeauthoryear {%
Zheng%
\ \protect \BOthers {.}}{%
Zheng%
\ \protect \BOthers {.}}{%
{\protect \APACyear {2021}}%
}]{%
zheng2021estimating}
\APACinsertmetastar {%
zheng2021estimating}%
\begin{APACrefauthors}%
Zheng, Z.%
, Curtis, J\BPBI H.%
, Yao, Y.%
, Gasparik, J\BPBI T.%
, Anantharaj, V\BPBI G.%
, Zhao, L.%
\BDBL {}Riemer, N.%
\end{APACrefauthors}%
\unskip\
\newblock
\APACrefYearMonthDay{2021}{}{}.
\newblock
{\BBOQ}\APACrefatitle {Estimating submicron aerosol mixing state at the global
  scale with machine learning and Earth system modeling} {Estimating submicron
  aerosol mixing state at the global scale with machine learning and earth
  system modeling}.{\BBCQ}
\newblock
\APACjournalVolNumPages{Earth and Space Science}{8}{2}{e2020EA001500}.
\PrintBackRefs{\CurrentBib}

\end{thebibliography}
\end{document}